\documentclass[
reprint,
11pt,
authoryear,
colorlinks=true,linkcolor=black,citecolor=black,urlcolor=black]{elsarticle}

\usepackage{isomath}
\usepackage{amsmath,amsthm}
\usepackage{amsbsy}
\usepackage{amssymb}
\usepackage{amscd}
\usepackage{amsfonts}
\usepackage{stmaryrd}
\usepackage{siunitx}
\usepackage{euscript}
\usepackage[utf8]{inputenc}
\usepackage[T1]{fontenc}
\usepackage{newtxtext} 
\everymath{\displaystyle}
\usepackage{exscale}

\usepackage{graphicx}
\usepackage{boxedminipage}
\usepackage{calc}
\usepackage[dvipsnames]{xcolor}
\graphicspath{ {media/} }
\usepackage[caption=false,justification=centerlast]{subfig}

\usepackage{setspace}
\usepackage{enumitem}
\setitemize{noitemsep,topsep=0pt,parsep=0pt,partopsep=0pt}
\setenumerate{noitemsep,topsep=0pt,parsep=0pt,partopsep=0pt}
\setdescription{noitemsep,topsep=0pt,parsep=0pt,partopsep=0pt}

\usepackage[colorinlistoftodos, color=green!40,prependcaption]{todonotes}

\usepackage{soul} 

\usepackage[normalem]{ulem}

\usepackage[small]{titlesec}

\titlespacing*{\section}{0pt}{12pt plus 4pt minus 2pt}{2pt plus 2pt minus 2pt}
\titlespacing*{\subsection}{0pt}{12pt plus 4pt minus 2pt}{2pt plus 2pt minus 2pt}
\titlespacing*\subsubsection{0pt}{12pt plus 4pt minus 2pt}{2pt plus 2pt minus 2pt}
\titlespacing*\paragraph{0pt}{12pt plus 4pt minus 2pt}{2pt plus 2pt minus 2pt}

\makeatletter

    \renewcommand*{\p@subsection}{}
    
    \renewcommand*{\p@subsubsection}{}
\makeatother

\makeatletter
\renewcommand\paragraph{\@startsection{paragraph}{4}{\z@}%
  {-3.25ex \@plus -1ex \@minus -.2ex}%
  {1em}%
  {\normalfont\normalsize\bfseries}}
\makeatother

\setuptodonotes{inline}
\usepackage{isomath}
\usepackage{amsmath}
\usepackage{amssymb}
\usepackage{amscd}
\usepackage{amsfonts}

\newcommand{\calE}{{\cal E}}

\newcommand{\calL}{{\cal L}}

\newcommand{\calY}{{\cal Y}}

\newcommand{\bfdelta}{\mathbold {\delta}}

\newcommand{\bfepsilon}{\mathbold {\epsilon}}

\newcommand{\bfmu}{\mathbold {\mu}}

\DeclareMathOperator{\trace}{tr}

\newcommand{\parderivsec}[2]{\frac{\partial^2 #1}{{\partial #2}^2}}

\newcommand{\parderiv}[2]{\frac{\partial #1}{\partial #2}}

\newcommand{\bfd}{{\mathbold d}}
\newcommand{\bfe}{{\mathbold e}}

\newcommand{\bfn}{{\mathbold n}}

\newcommand{\bfx}{{\mathbold x}}

\newcommand{\bfA}{{\mathbold A}}

\newcommand{\bfD}{{\mathbold D}}

\newcommand{\bfI}{{\mathbold I}}

\newcommand{\bfL}{{\mathbold L}}
\newcommand{\bfM}{{\mathbold M}}

\newcommand{\bfR}{{\mathbold R}}

\newcommand{\bfT}{{\mathbold T}}

\newcommand{\bfY}{{\mathbold Y}}

\usepackage{microtype}

\usepackage{setspace}
\usepackage[left=1in,right=1in,top=1in,bottom=1in]{geometry}

\usepackage[mathlines]{lineno}

\usepackage{orcidlink}
\usepackage{algorithm}
\usepackage{algpseudocode}

\makeatletter
\def\ps@pprintTitle{%
 \let\@oddhead\@empty
 \let\@evenhead\@empty
 \let\@oddfoot\@empty
 \let\@evenfoot\@empty
}
\makeatother

\usepackage{esint}
\usepackage{tcolorbox}
\usepackage[title]{appendix}

\DeclareMathOperator{\IM}{Im}

\newcommand{\bfepseff}{{\bfepsilon_{*}}}
\newcommand{\epseff}{{\epsilon_{*}}}
\newcommand{\sfred}[1]{\textcolor{red}{#1}}

\newcommand{\bfcalL}{\mathbf{\calL}}

\newcommand{\dprime}{{\prime\prime}}
 
\newcommand{\epsOne}{{\bfepsilon_{1}}}
\newcommand{\epsTwo}{{\bfepsilon_{2}}}
\newcommand{\epsOneReal}{{\bfepsilon_{1}^\prime}}
\newcommand{\epsOneIm}{{\bfepsilon_{1}^{\prime\prime}}}
\newcommand{\epsTwoReal}{{\bfepsilon_{2}^\prime}}
\newcommand{\epsTwoIm}{{\bfepsilon_{2}^{\prime\prime}}}

\AtBeginDocument{\hypersetup{linkcolor=red}}
\begin{document}

\begin{frontmatter}



\title{\Large{Bounds on the Uniaxial Effective Complex Permittivity of Two-phase Composites and Optimal or Near Optimal Microstructures}} 

\author[label1]{Kshiteej J. Deshmukh\orcidlink
{0000-0002-6825-4280}\corref{cor1}\fnref{label2}}
\ead{kdeshmukh@uh.edu}
\cortext[cor1]{Corresponding author}
\fntext[label2]{Now at, Department of Mechanical and Aerospace Engineering, University of Houston, Houston, 77204, TX, USA}

\author[label1]{Graeme W. Milton\orcidlink{0000-0002-4000-3375}} 
\ead{graeme.milton@utah.edu}

\affiliation[label1]{organization={Department of Mathematics, University of Utah},
            city={Salt Lake City},
            postcode={84101},
            state={UT},
            country={USA}}



\begin{abstract}
Electromagnetic materials with a uniaxial effective permittivity tensor, characterized by its transverse ($\epsilon_\perp$) and axial ($\epsilon_\parallel$) components, play a central role in the design of advanced photonic and electromagnetic materials  including hyperbolic metamaterials, and biological imaging platforms. 
    Tight bounds on the complex effective permittivity of such metamaterials are critical for predicting and optimizing their macroscopic electromagnetic response.
    While rigorous tight bounds exist for isotropic two-phase composites, corresponding results for uniaxial composites remain relatively unexplored.
    In this work, we systematically investigate the attainable range of $\epsilon_\perp$ and $\epsilon_\parallel$ in the quasistatic regime for two-phase metamaterials with isotropic homogeneous phases. 
    By analyzing known microgeometries and constructing hierarchical laminates (HLs), we demonstrate that the classical bounds on $\epsilon_\perp$ are not optimal. 
    We conjecture improved bounds based on numerically fitted circular arcs derived from convex hulls of $\epsilon_\perp$ values obtained from HLs, and we identify optimal rank-4 HL structures that achieve all points on the conjectured bounds. 
    Additionally, we quantify the correlation between $\epsilon_\perp$ and $\epsilon_\parallel$ for fixed volume fractions, and propose a design algorithm to construct HL microstructures achieving prescribed values of $\epsilon_\perp$. 
    Leveraging the Cherkaev-Gibiansky transformation and the  translation method, we extend recent techniques developed for isotropic composites by Kern-Miller-Milton to derive translation bounds on the uniaxial complex effective permittivity tensor. 
    Finally, bounds on the sensitivity of the effective permittivity tensor of low-loss composites are obtained and their optimality is shown in two-dimensions.
    Our results advance the theoretical understanding of uniaxial metamaterials and provide practical tools for the design of tailored anisotropic metamaterials.
\end{abstract}



\begin{keyword}
Homogenization \sep 
Two-phase composites \sep 
Hyperbolic metamaterials\sep 
Uniaxial complex effective permittivity \sep



\end{keyword}

\end{frontmatter}

\section{Introduction}\label{Introduction}
A material with a uniaxial permittivity tensor represented by the matrix $\bfepsilon$ is characterized by two of its eigenvalues in the directions orthogonal to the symmetry-axis being equal (transverse component denoted by $\epsilon_\perp$), and the eigenvalue along the axis of symmetry is the axial component $\epsilon_\parallel$, for instance,
\begin{equation}
    \bfepsilon = \begin{bmatrix}
        \epsilon_\perp &0&0\\
        0&\epsilon_\perp&0\\
        0&0&\epsilon_\parallel
    \end{bmatrix}.
\end{equation}
Uniaxial materials can be found in nature, e.g. calcite and quartz crystals, biological tissues like lipid bilayers, as well as can be artificially constructed, for example by taking a stack of sheets of two different isotropic dielectric materials.
The study of effective uniaxial permittivity tensors of uniaxial dielectric composites has become increasingly important in electromagnetic applications.
Recently, a computational microscopy method was used to determine the components of the uniaxial permittivity tensor and uncover the biological architecture of mouse brain tissue  \citep{yeh2021upti,yeh2024permittivity}.
Hyperbolic metamaterials  \citep{jacob2006optical,salandrino2006far,poddubny2013hyperbolic,shekhar2014hyperbolic} (also known as HMMs) have gained a widespread interest due to their ability to resolve sub-wavelength objects as they can support propagation of large wavevectors.
They are being used in several exotic applications like, optical negative refraction \citep{smith2004negative,hoffman2007negative}, spontaneous emission engineering \citep{noginov2010controlling,lu2014enhancing}, and optical hyperlenses for sub-diffraction limit imaging \citep{jacob2006optical,liu2007far}.
HMMs represent a special class of dielectric metamaterials with uniaxial effective  permittivity tensors where ideally $\epsilon_\perp$ and $\epsilon_\parallel$ are real and  either $\epsilon_\perp<0$ and $\epsilon_\parallel>0$, or $\epsilon_\perp>0$ and $\epsilon_\parallel<0$.
In practice, while one can have real positive modulus, the other modulus is complex with a negative real part and a small imaginary part.
In this work, we focus on the more general quasistatic two-phase metamaterials with complex uniaxial effective permittivity tensors.

The study of effective permittivity in composite materials is a fundamental problem in electromagnetism, with significant implications for the design of dielectric materials, metamaterials, and photonic structures. Determining accurate bounds on the complex permittivity of such composites is essential for characterizing their macroscopic electromagnetic response. In this context,  \cite{schulgasser1976bounds} analyzed a statistically homogeneous and isotropic two-phase composite where the constituent phases were low-loss dielectrics. 
Under the assumption that the imaginary part of the permittivity of each phase is small compared to the real part, they derived bounds on the effective complex permittivity of the composite. 
For isotropic two-phase composites with unknown or known volume fractions bounds on complex effective permittivity are given by the well-known Bergman-Milton \citep{milton1980bounds,PhysRevLett.45.148.2,milton1981bounds,milton1981bounds-b} bounds, which correspond to circular arcs in the complex plane.
Recently,  \cite{kern2020tight} improved one of the circular arcs bounding the complex isotropic permittivity and showed that it is optimal, as it is attained by assemblages of doubly coated cylinders.
Bounds on the transverse and axial components of the uniaxial complex effective permittivity for a two-phase composite were given by \cite{milton1981bounds}, but the existence of optimal microstructures was not fully investigated, and the question of whether tighter bounds are possible remains unanswered.

In this work, we investigate the bounds and optimal microstructures pertaining to the problem of uniaxial effective permittivity of two-phase composites with isotropic homogeneous constituent phases in the quasistatic regime, i.e. under the assumption that the composite microscale variations occur at much smaller length scales than the wavelength of the fields.
The isotropic permittivity of the pure phases is denoted by $\epsilon_1$ and $\epsilon_2$, and their respective volume fractions are $f_1$ and $1-f_1$.
The complex effective permittivity tensor is denoted by $\bfepseff$.
Section \ref{sec:Loci} provides a brief background on the Bergman-Milton bounds.    
In Sections \ref{sec:Loci} and \ref{sec:Effective Uniaxial Permittivity From Known Microgeometries}, we investigate the range of $\epsilon_\perp$ obtained by considering well-known simple geometries and hierarchical laminates (HLs).
The results show that one of the bounds for the uniaxial case given by \cite{milton1981bounds} is likely not optimal. 
By numerically fitting circles to the convex hull of $\epsilon_\perp$ values obtained from Schulgasser laminates of hierarchical laminates we conjecture optimal bounds given by the fitted circular arcs. 
Tree structures of the HLs attaining points on the conjectured bounds are identified and they are rank-4 HLs. 
Laminates formed in more than one lamination step are called hierarchical laminates and the minimum number of sufficiently separated length scales in the subsequent lamination steps is defined as the rank of the hierarchical laminate.

From the perspective of designing uniaxial metamaterials it is highly desirable to know if $\epsilon_\perp$ or $\epsilon_\parallel$, or both can be customized. Towards this end, in Section \ref{sec:correlation} we give the correlation between the axial and the transverse components of a given two-phase composite with fixed volume fraction. 
We also propose an algorithm to design a microstructure (specifically, a HL geometry) that achieves the exact specified value of  $\epsilon_\perp$ if it lies in the interior of the region enclosed by the bounds.

Using the Cherkaev-Gibiansky transformation \citep{cherkaev1994variational} and the translation method  \citep{Tartar1985,murat_tartar_1985,lurie1982conductivity,lurie1984exact,milton2002theory},  \cite{kern2020tight} proposed optimal bounds for the complex isotropic effective permittivity. Following their analysis we use the same translation tensor to find bounds for uniaxial complex effective permittivity. The corresponding results are given in Section \ref{sec:tight}. In Section \ref{app:A}, we obtain bounds on the sensitivity of the anisotropic effective permittivity in low-loss composites in two-dimensions.

All numerical calculations are performed using Python.  The values of $\epsilon_1$,  $\epsilon_2$, $f_1$ for all the plots in the paper are chosen to clearly illustrate the findings. Below is a list of common notations that will be used throughout this paper.
\paragraph*{Notation}
\begin{enumerate}
    \item Electric displacement field - $\bfd(\bfx)$
    \item Electric field - $\bfe(\bfx)$
    \item Complex permittivity tensor - $\bfepsilon(\bfx)$
    \item Complex effective permittivity tensor- $\bfepseff$
    \item Permittivity of phase-1 and phase-2 - $\epsilon_1$, $\epsilon_2$, respectively.
    \item Volume fractions of phase-1 and phase-2 - $f_1$ and $f_2$, respectively.
    \item Transverse component of $\bfepseff$ and axial component of  $\bfepseff$ - $\epsilon_\perp$ and  $\epsilon_\parallel$, respectively
    \item Microgeometries are denoted as:
    \\
    CC1 -- Coated cylinder with phase 1 as core \\
    CC2 -- Coated cylinder with phase 2 as core \\
    CE1 -- Coated ellipsoid with phase 1 as core \\
    CE2 -- Coated ellipsoid with phase 2 as core \\
    CEC1 -- Coated elliptical cylinder with phase 1 as core \\
    CEC2 -- Coated elliptical cylinder with phase 2 as core \\
    COS1 -- Coated oblate spheroid with phase 1 as core \\
    COS2 -- Coated oblate spheroid with phase 2 as core \\
    CPS1 -- Coated prolate spheroid with phase 1 as core \\
    CPS2 -- Coated prolate spheroid with phase 2 as core
    \\
     DCC1 -- Doubly coated cylinder with phase 1 as core \\
    DCC2 -- Doubly coated cylinder with phase 2 as core \\
    HS1 -- Hashin-Shtrikman coated sphere assemblage with phase 1 as core \\
    HS2 -- Hashin-Shtrikman coated sphere assemblage with phase 2 as core \\
    L -- Simple laminates \\
    HL -- Hierarchical laminates
\end{enumerate}

\section{Range of \texorpdfstring{$\epsilon_\perp$ for a Two-phase Composite with Arbitrary Volume Fraction}{TEXT}}\label{sec:Loci}
\begin{figure}[t!]
    \centering
    \includegraphics[width=0.8\textwidth]{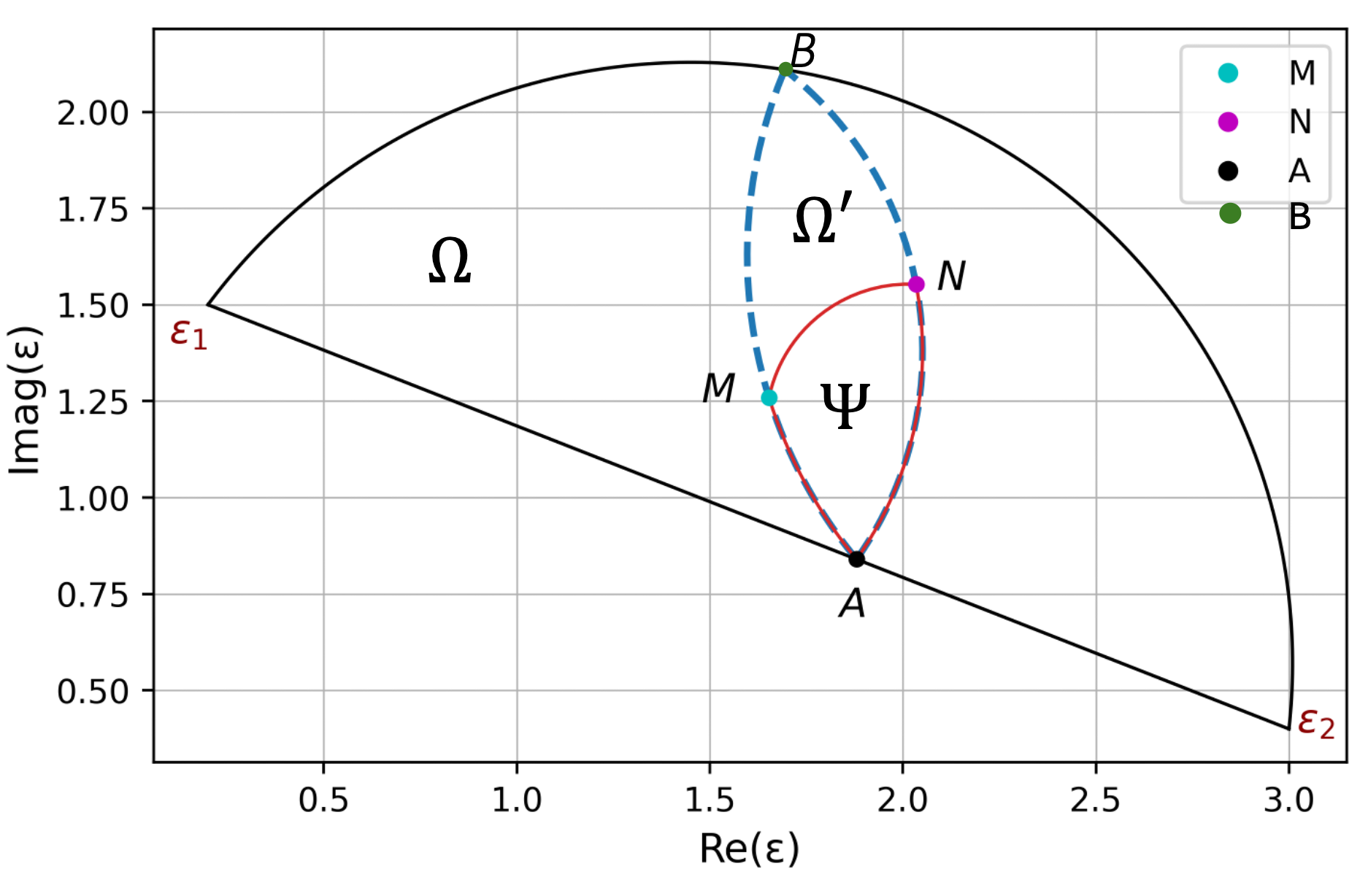}
    \caption{Bounds on the complex effective permittivity of a two-component composite with $\epsilon_1=0.2+1.5 \imath$, $\epsilon_2=3+0.4\imath$, and  $f_1=0.4$.
    The region $\Omega$ gives bounds on the effective permittivity when the volume fraction is unknown.
    Bounds on the transverse component $\epsilon_\perp$ of the uniaxial permittivity are shown by the solid red curves enclosing the region $\Psi$.
    The region $\Psi$ is bounded by three circular arcs $AM$, $MN$, and $AN$. 
    The axial component can lie anywhere in the region $\Omega^\prime$ (shown by the dashed blue curve)  bounded by the circular arcs $AMB$ and $ANB$.
    }
    \label{fig:Bounds Milton1981}
\end{figure}
Around 1980s, Bergman and Milton \citep{PhysRevLett.45.148.2,milton1980bounds,milton1981bounds,milton1981bounds-b} derived bounds on the complex effective permittivity of  a two-phase composite in the quasistatic limit.
Depending on the known information of the composite these bounds were shown to be regions in the complex plane bounded by circular arcs.
For a two-phase composite with isotropic and homogeneous component complex permittivities given by $\epsilon_1$ and $\epsilon_2$, these bounds are summarized in Figure \ref{fig:Bounds Milton1981} as follows.
If only the phase permittivities $(\epsilon_1,\epsilon_2)$ are known then the effective permittivity $\epseff$, taken to be any diagonal element of the complex effective permittivity matrix $\bfepseff$, is restricted to the region $\Omega$ bounded between a straight line joining $\epsilon_1$ and $\epsilon_2$, and a circular arc passing through $\epsilon_1, \epsilon_2$ and the origin.
If the volume fractions $f_1$ and $1-f_1$ of the two phases are known, then $\epseff$ is bounded by the lens shaped region $\Omega^\prime$ (shown by dashed blue curves in Figure \ref{fig:Bounds Milton1981}) bounded by two circular arcs $AMB$ and $ANB$. 
The circular arc $AMB$ is generated by the circle passing through the points $A$, $B$, $\epsilon_2$, and the arc $ANB$ is generated by the circle passing through points $A$, $B$, $\epsilon_1$.
For uniaxial composite materials, the effective permittivity has two of the three eigenvalues equal to each other, which we refer to as the transverse component of $\bfepseff$, and the remaining eigenvalue is the axial component.
The axial component, $\epsilon_\parallel$, can take any value in the region $\Omega^\prime$, but the transverse component, $\epsilon_\perp$, is restricted to the region $\Psi$ contained within $\Omega^\prime$ and bounded by circular arcs $AM$, $MN$, and $AN$ (colored in red in Figure \ref{fig:Bounds Milton1981}).
Since the focus of this paper is on effective complex uniaxial permittivities, we will not discuss the further bounds obtained in the isotropic case.
\begin{figure}[t!]
    \centering
    \includegraphics[width=0.9\textwidth]{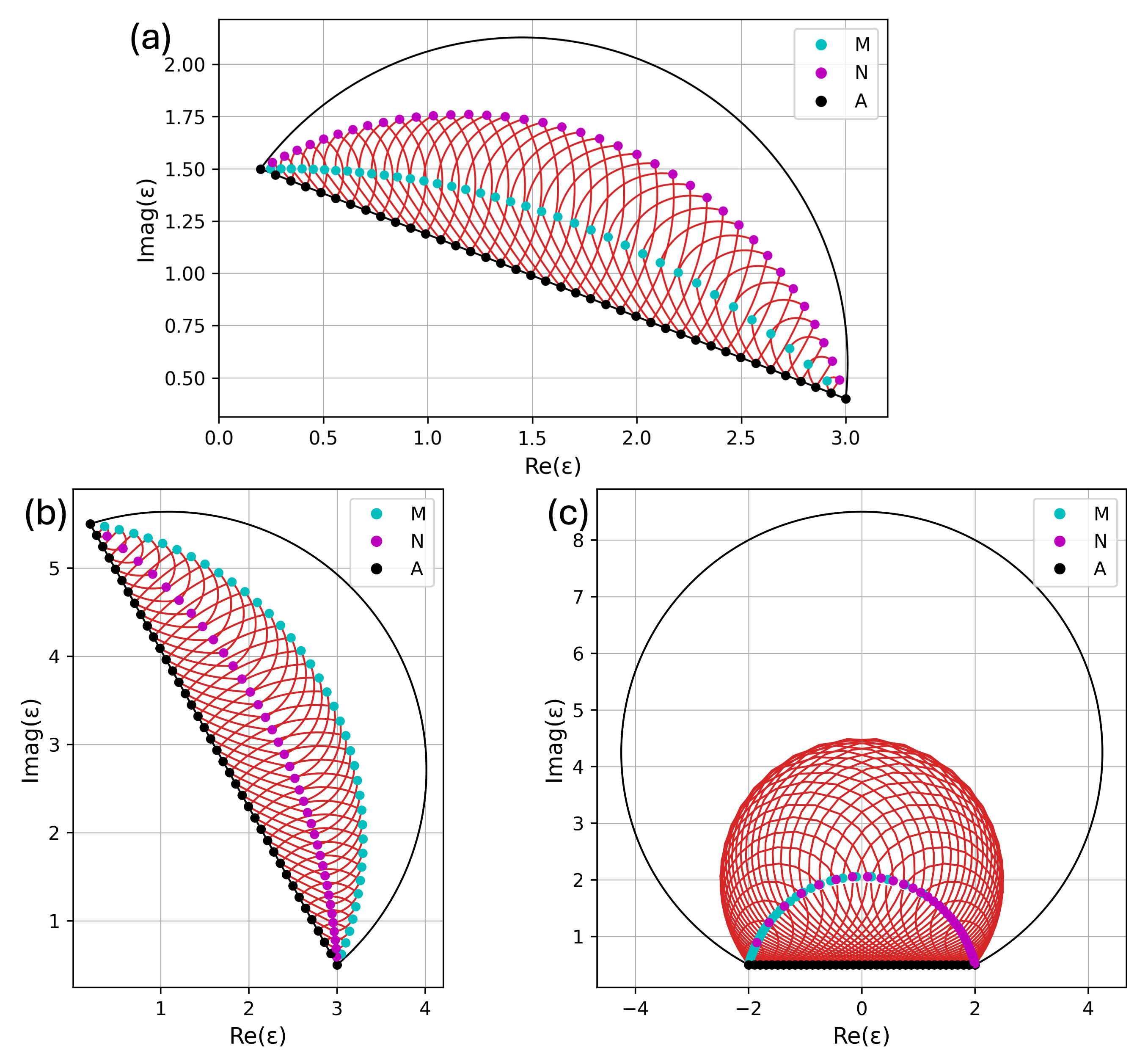}
    \caption{Bounds obtained from the union of $\Psi$ regions by varying the volume fraction $f_1$.
    Different values for $\epsilon_1$ and $\epsilon_2$ result in different parts of the region $\Psi$ forming the bound.
    (a) $\epsilon_1=0.2+1.5\imath $, $\epsilon_2=3+0.4\imath$: Loci of $N$ forms the bound. 
    (b) $\epsilon_1 = 0.2 + 5.5\imath $, $\epsilon_2=3 + 0.5\imath$: Loci of  $M$ forms the bound. 
    (c) $\epsilon_1 = -2 + 0.5\imath $, $\epsilon_2=2 + 0.5\imath$: Points on the boundary $MN$ form the bound.
    Points $N$, $M$, and $A$ are shown in pink, cyan, and black, respectively.
    }
    \label{fig:Locii}
\end{figure}
In this section, we are interested in determining all the possible values attained by the transverse component $\epsilon_\perp$, i.e., the range of $\epsilon_\perp$, for a composite material with known phase permittivities but arbitrary volume fractions of the phases. 
A simple way to do that is to look at the bounds formed by the union of the regions $\Psi$ as the volume fraction $f_1$ is varied from $0$ to $1$ for given values of $(\epsilon_1, \epsilon_2)$.
Figure \ref{fig:Locii} shows the range of $\epsilon_\perp$ generated by plotting the $\Psi$ regions as the volume fraction $f_1$ is varied. 
Depending on the phase permittivities of the composite, the outer bound may be formed by the loci of points $N$ (purple), loci of points $M$ (cyan), or by points on the arc $MN$ (red curves) as seen in Figures \ref{fig:Locii}\textcolor{red}{(a)-(c)}, respectively.
The other part of boundary of the union of the $\Psi$ regions in each case is formed by the loci of points $A$ (black).


\section{Effective Uniaxial Permittivities From Known Microgeometries}\label{sec:Effective Uniaxial Permittivity From Known Microgeometries}
\begin{figure}
    \centering
    \includegraphics[width =\linewidth]{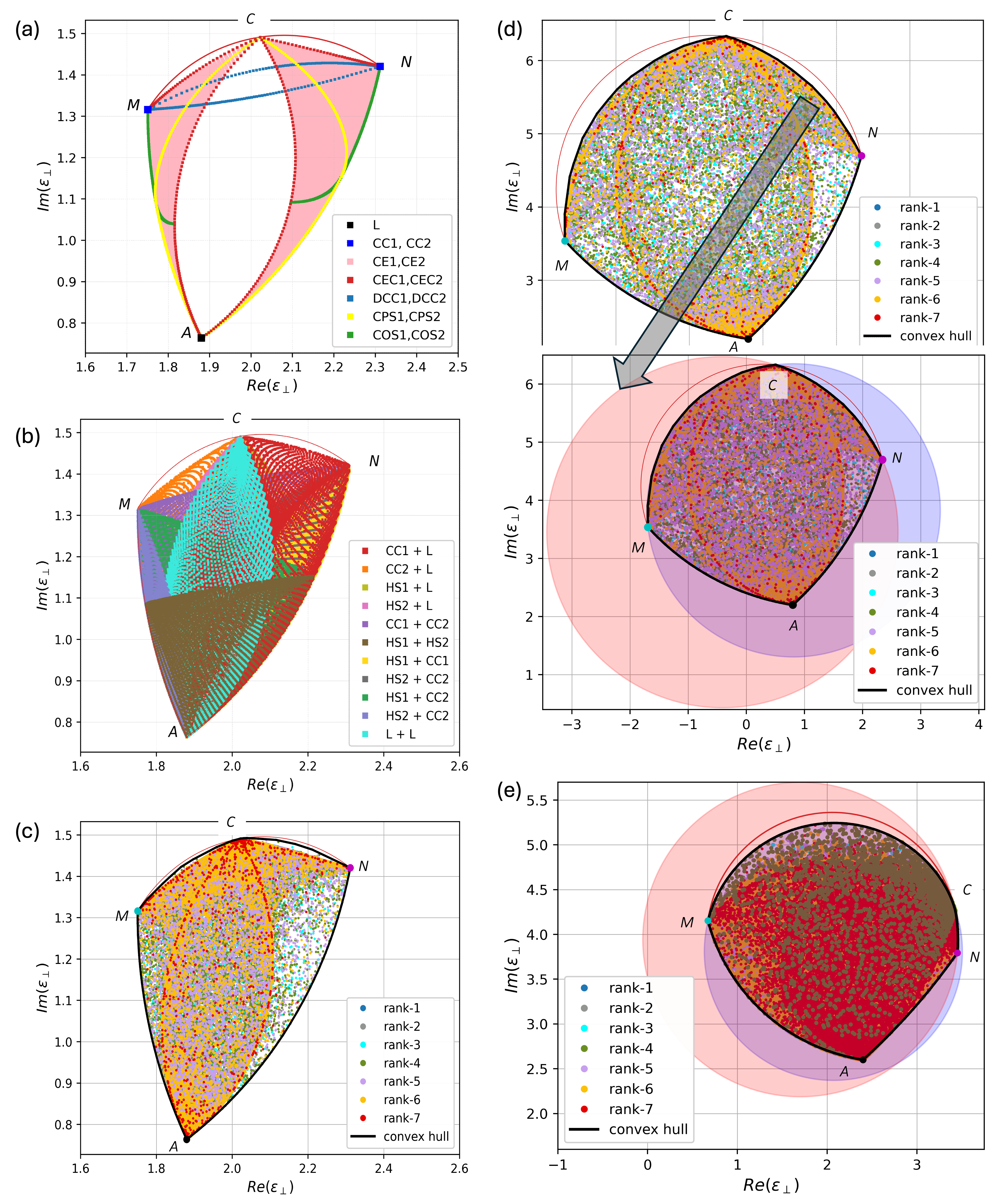}
    \caption{Microgeometries filling the  region $\Psi$ and conjectured optimal bounds.
    (a)-(c) show results for a composite with $\epsilon_1=0.2+1.76\imath$, $\epsilon_2=3+0.1\imath$, and $f_1=0.4$.
    The ranges of $\epsilon_\perp$ (shown by colored markers) are obtained from Schulgasser laminates of: 
    (a) assemblages of well-known microgeometries where points $A$, $M$, and $N$ are attained by the L, CC1, and CC2 geometries respectively,
    (b)  assemblages of mixed microgeometries,
    (c) HL geometries. The convex hull of range of $\epsilon_\perp$ is shown by the solid black curve.
    (d)-(e) show results for a composite with $\epsilon_1=-4+1\imath$, $\epsilon_2=4+3\imath$, and $f_1=0.4$. and $f_1=0.2$, respectively.
    The red circle is fitted to the points on the boundary $CN$ of the convex hull (the grey arrow in (d) points to the red circle that is fitted to points on the boundary $CN$), and similarly the blue circle is fitted to points on the boundary $MC$ of the convex hull.}
    \label{fig:Psi_fill}
\end{figure}When dealing with bounds on effective properties, it is natural and extremely desirable to find the optimal microgeometries.
While the bounds on uniaxial permittivity by \cite{milton1981bounds} (discussed above) were given more than four decades ago, the uniaxial microgeometries that fill up the region $\Psi$ have not been explored.
Using the Schulgasser lamination technique \citep{schulgasser1977bounds}, one can  construct uniaxial effective permittivity tensors from composites with diagonal permittivity tensors.
For such a material, the permittivity can be represented by $\bfepsilon = \mathrm{Diag}[\lambda_1,\lambda_2, \lambda_3]$, where $\lambda_i$ ($i=1,2,3$) denote the principal permittivities.
Previously, Schulgasser laminates were constructed to obtain hierarchical laminates with an isotropic effective permittivity given by $\trace{(\bfepseff)}/3$ from an anisotropic material  \citep{schulgasser1977bounds,kern2020tight}.
Here, we use a similar Schulgasser lamination scheme, but stop  after a single lamination step (such a scheme was used in \cite{milton1981bounds}).
Suppose we construct a composite by laminating this anisotropic material with volume fraction $f$ along the direction $\bfn$ parallel to the $x_1$-axis (principal direction corresponding to $\lambda_1$) with a $90^\circ$ rotation of itself about the $x_1$-axis with permittivity tensor $\bfepsilon_0 = \mathrm{Diag}[\lambda_1,\lambda_3, \lambda_2]$ and volume fraction $1-f$.
Then the effective permittivity of this laminate is simply given by taking the arithmetic and harmonic means as, 
\begin{equation}\label{Schulgasser step}
    \bfepseff 
    = 
    \begin{bmatrix}
        \lambda_1 & 0 &0\\
        0& f\lambda_2+(1-f)\lambda_3 &0\\
        0& 0& f\lambda_3+(1-f)\lambda_2
    \end{bmatrix}
\end{equation}
When $f=0.5$, we get the uniaxial permittivity tensor $\bfepseff=\mathrm{Diag}[\lambda_1,(\lambda_2+\lambda_3)/2 , (\lambda_3+\lambda_2)/2]$.
Hereafter in this article, we consider laminates with only orthogonal directions of lamination, and we refer to the uniaxial laminates created by this scheme in \eqref{Schulgasser step} as the Schulgasser laminates.
For a given anisotropic tensor $\bfepsilon$ with $\lambda_1\neq\lambda_2\neq\lambda_3$, one can generate three different uniaxial laminates by constructing Schulgasser laminates along each of the three principal directions.
This implies that for a given anisotropic tensor $\bfepsilon = \mathrm{Diag}[\lambda_1,\lambda_2, \lambda_3]$, we can construct three uniaxial laminates with the transverse components given by $(\lambda_2+\lambda_3)/2$, $(\lambda_1+\lambda_3)/2$, $(\lambda_1+\lambda_2)/2$, and their respective axial components as $\lambda_1$, $\lambda_2$, $\lambda_3$.

We first investigate the range of $\epsilon_\perp$ of the Schulgasser laminates constructed from well-known microgeometries, and then identify the optimal microstructures and also those that fill the region $\Psi$. 
The effective permittivity $\bfepseff$ of an assemblage of confocal coated ellipsoids (CE) with a core of phase 1 can be obtained from the relation  \citep{milton1981bounds,milton2002theory,tartar1985estimations}:
\begin{equation}
    f_1\epsilon_2 (\bfepseff-\epsilon_2\bfI)^{-1}
    =
    \epsilon_2(\epsilon_1-\epsilon_2)^{-1}\bfI
     + f_2 \bfM,
\end{equation}
where $\bfM$ is related to the depolarization tensors $\bfD_c$ and $\bfD_e$ of the core and the exterior elliptical surfaces respectively, by the relation $\bfM = (\bfD_c - f_1\bfD_e)/f_2$.
It was shown that  $\trace{\bfM}=1$, and that $\bfM$ is positive semi-definite.
As the shape of the coated ellipsoids is varied keeping the volume fraction $f_1$ fixed, $\bfM$ varies over all positive semi-definite symmetric matrices satisfying $\trace{\bfM}=1$.
As the coated ellipsoids are confocal the depolarization tensors $\bfD_c$ and $\bfD_e$ are not independent of each other. 
If the principal axes of the ellipsoids are chosen as the coordinate system, then $\bfM$, $\bfD_c$, $\bfD_e$ are diagonal matrices.
Special cases of the coated ellipsoid geometry can be obtained depending on the depolarization factors.
For coated spheroids, the tensor $\bfM=\mathrm{Diag}[1-2m,m,m]$ with $m\in[0,1/2]$ as two of the depolarization factors are equal.
They can be further distinguished into coated prolate and coated oblate spheroids with the same tensor $\bfM=\mathrm{Diag}[1-2m,m,m]$  but with  $m\in[0,1/3]$ and $m\in[1/3,1/2]$, respectively.
For coated elliptical cylinders (CEC), one of the depolarization factors is $0$ and we get the tensor $\bfM=\mathrm{Diag}[m,1-m,0]$ with $m\in[0,1]$.

Figure \ref{fig:Psi_fill}\sfred{(a)} shows the range of $\epsilon_\perp$ obtained from the Schulgasser laminates of assemblages of CEs and assemblages of doubly coated cylinders (DCC).
The parts of the boundary $AM$ and $AN$ are attained using coated oblate and prolate spheroids (shown in green and yellow, respectively) as noted in \cite{bergman1980exactly}.
The boundary of $\Omega^\prime$ is also attained by coated elliptical cylinders as noted in \cite{milton1981bounds}.
Schulgasser laminates of assemblages of CEC (shown in red) achieve only points $M$, $N$, and $C$ on the boundary $MCN$.
At the point $C$, the CEC geometry degenerates to a simple laminate geometry.
Among the already uniaxial geometries,  point $A$ is attained by simple laminate (L) geometry, whereas the points $M$ and $N$ are obtained from assemblages of coated cylinders with phase-1 as core (CC1) and phase-2 as core (CC2) respectively.
The transverse component for an assemblage of CC1 geometry is given by the relation \citep{hashin1962elastic,hashin1962variational},
\begin{equation}\label{CC1 trans}
    {\epsilon_\perp} = 
    \epsilon_2
    +
    \frac{2f_1\epsilon_2(\epsilon_1-\epsilon_2)}{2\epsilon_2+f_2(\epsilon_1-\epsilon_2)}.
\end{equation}
The transverse component of assemblages of the DCC1 geometry with the core and outermost cylinders of phase-1 with volume fractions $p_1$ and $p_3$ and the middle cylindrical shell of phase-2 with volume fraction $p_2=1-p_1-p_3$ can be given as \citep{schulgasser1977concerning}, 
\begin{equation}
    \epsilon_\perp =
    \epsilon_1 + \frac{2(p_1+p_2)\epsilon_1}{p_3-\frac{2\epsilon_1}{\epsilon_1 - \epsilon_0}},
\end{equation}
where
\begin{equation}
    \epsilon_0 = \epsilon_2 + 
    \frac{2p_1/(1-p_3)\epsilon_2}{p_2/p_3 -2\frac{\epsilon_2}{\epsilon_2-\epsilon_1}}.
\end{equation}
Analogous expressions hold for the DCC2 geometry with a core of phase 2.
By varying the volume fractions $p_1$ and  $p_3$ of the core and outer cylinders (phase-1) while keeping the volume fraction $p_2=1-f_1$ (phase-2) fixed, we get the blue curve in Figure \ref{fig:Psi_fill}\sfred{(a)}  showing the range of transverse components of DCC1 and DCC2 assemblages.

Laminate geometries can be constructed from assemblages of different geometries; for instance,  assuming sufficient separation of scales, an assemblage of CC can be laminated with an assemblage of coated spheres to get an effective permittivity tensor (which maybe anisotropic), and further Schulgasser laminates of these structures can be constructed to obtain uniaxial laminates.
Figure \ref{fig:Psi_fill}\sfred{(b)} shows the range of $\epsilon_\perp$ of Schulgasser laminates of such mixed geometries.
Schulgasser laminates of a simple laminate (L) with a $90^\circ$ rotation of itself yields the point $C$ on the boundary $MN$ of the region $\Psi$ with the transverse component given by,
\begin{equation}
    \epsilon_\perp = 
    \frac{1}{2}
    \left(f_1\epsilon_1 + f_2\epsilon_2 + 
    \frac{1}{f_1/\epsilon_1 + f_2/\epsilon_2}\right).  
\end{equation}

Both Figures \ref{fig:Psi_fill}\sfred{(a)-(b)} show that the bounds $AM$ and $AN$ are optimal as those can be attained by coated oblate and prolate spheroids, and  mixtures of coated cylinder and laminate geometries. 
However, only three points on the arc $MCN$ have been attained.

\subsection{Conjectured Bounds on the Transverse Component of theUniaxial Permittivity}
The laminates constructed by laminating two phases in more than one step (assuming sufficient separation of scales at every subsequent step) are called hierarchical laminates (HL).
The minimum number of  length scales required for their construction is typically referred to as the rank of the hierarchical laminate.
The simple laminate geometry L is a rank-1 laminate structure formed by laminating two pure phases in one step. 
For a rank-2 laminate, the rank-1 laminate (core phase)  is further laminated with one of the pure phases or another rank-1 laminate  (coating phase).
In subsequent lamination steps, the laminate generated in the previous step is laminated with one of the pure phases or any one of the previously generated laminates, thereby generating higher ranked laminates. 
In this paper, we consider HLs where subsequent lamination directions are orthogonal to each other, and hence there are only three possible lamination directions, i.e. along the $\bfx_1,\bfx_2$, and $\bfx_3$ axes. 
Finally, to obtain uniaxial geometries we consider Schulgasser laminates of the HL, thereby adding a lamination step and increasing the rank by one of the given HL.
The other geometries can be seen as special cases of the HL geometries since CC1 and CC2 can be replaced by HL (as can all the geometries listed in \sfred{8} of the notation section). 

For given $\epsilon_1,\epsilon_2$ values of the pure phase with a fixed volume fraction $f_1$, we numerically construct about a million HLs ranging from rank-1 to rank-8.
The range of $\epsilon_\perp$ of the Schulgasser laminates of these HLs is shown in Figure \ref{fig:Psi_fill}\sfred{(c)}. 
It is observed that HLs can densely populate the region $\Psi$ and attain  points everywhere, except close to the arc $MCN$.
A convex hull of all the $\epsilon_\perp$ values obtained from the HLs is plotted and shown by the solid black curve in Figure \ref{fig:Psi_fill}.
The convex hull shows that optimal HL microstructures can attain all points on the black arcs $AN$ and $AM$, while the only points attained on the boundary $MCN$ in this case too are $M$, $C$, and $N$.
This observation is common to other composites considered in the analysis, for example, in Figure \ref{fig:Psi_fill}\sfred{(d)-(e)} we consider a composite with $\epsilon_1 = -4+1\imath$, $\epsilon_2=4+3\imath$, and $f_1=0.4$ (the values are arbitrarily chosen to clearly show the plots), and one can observe from the convex hull that only points $M$, $C$, and $N$ are attained on the boundary.
This suggests that the original bound corresponding to arc $MCN$ (shown by solid red curve) is not the best possible, and that tighter bounds may exist.

On further analysis of the convex hulls considered for several examples in our analysis, we find that the parts of the boundary $MC$ and $NC$ of the convex hull that we label $\Gamma$ can be fit to a good accuracy by two circular arcs; one passing through the points $M$ and $C$, and the other passing through the points $N$ and $C$, respectively. 
In Figure \ref{fig:Psi_fill}\sfred{(d)} the red shaded circle fits the boundary $NC$ of $\partial\Gamma$ and the blue shaded circle fits the boundary $MC$ of $\partial\Gamma$. 
Hence, we propose that the fitted circular arcs $MC$ and $NC$ on $\Gamma$ are the conjectured optimal bounds on $\epsilon_\perp$ obtained numerically.
In our numerical calculations, we observe that the fitted red and blue circles corresponding to the conjectured bounds do not pass through any obvious third point (even in the Y-transformed plane - see \cite{milton2002theory} for definition) that can completely characterize the circles.
We will refer to the fitted circular arcs $MC$ and $NC$ on $\Gamma$, as the blue circular arc $MC$ and the red circular arc $NC$, respectively.
Figure \ref{fig:Psi_fill}\sfred{(e)} shows the conjectured bounds given by the blue circular  arc $MC$ and red circular arc $NC$ on $\partial\Gamma$ for the same composite with $\epsilon_1 = -4+1\imath$, $\epsilon_2=4+3\imath$, but $f_1=0.2$.
These numerical results show that the original region $\Psi$ which was bounded by three circular arcs, is now bounded by four circular arcs.

\begin{figure}
    \centering
    \includegraphics[width=0.75\linewidth]{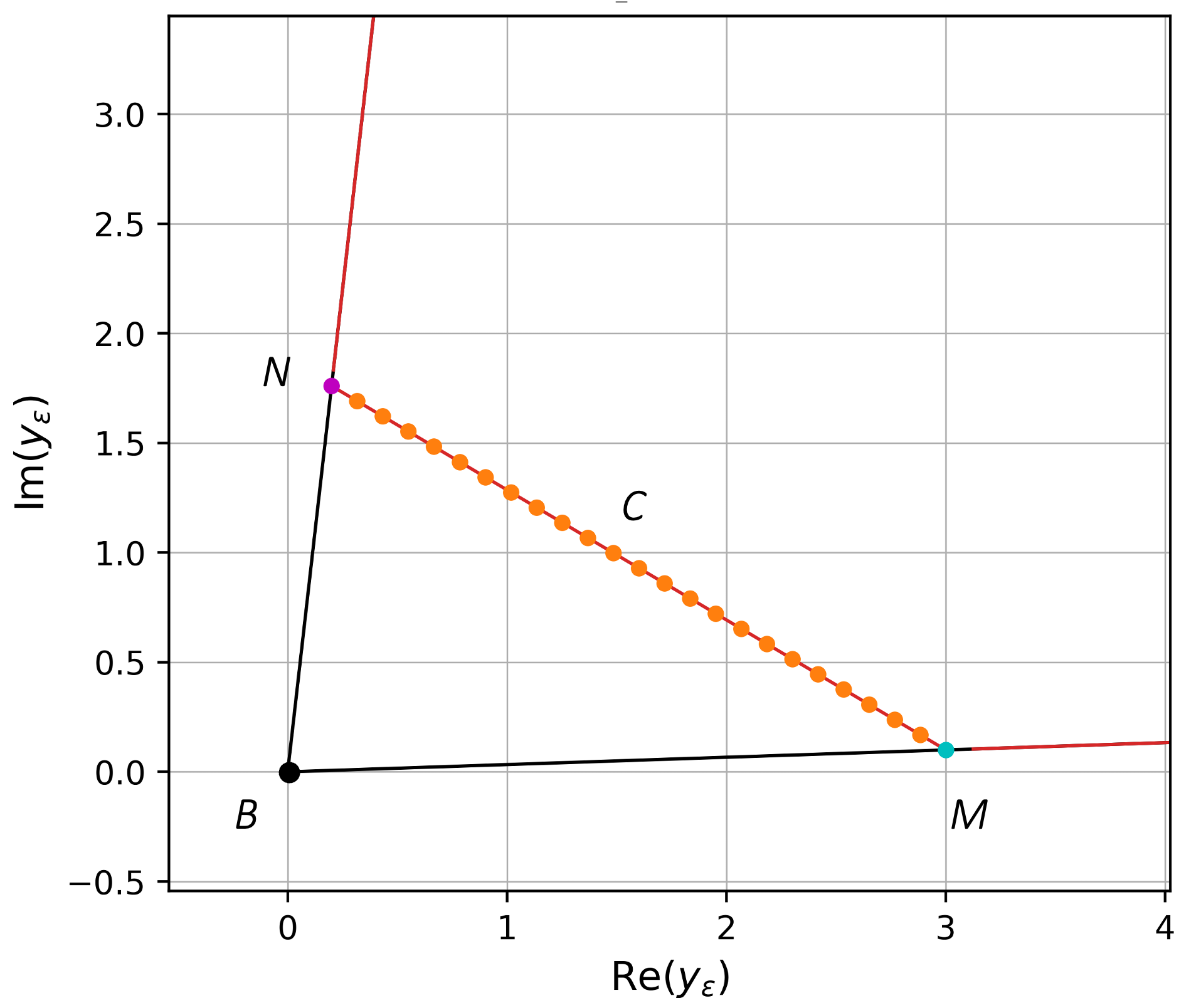}
    \caption{Y-transform of the region $\Psi$ (bounded by the red straight lines extending from points $M$ and $N$ to infinity, and the line segment $MN$) is shown with points $M, N, B$, and $C$ shown in cyan, pink, black, and orange, respectively, for composite with $\epsilon_1=0.2+1.76\imath$, $\epsilon_2=3+0.1\imath$.
    In the y-plane, the region $\Psi$ is volume fraction independent and the point $C$, attained by the Schulgasser laminate of L, traces the bound given by the straight line joining points $M$ and $N$ (which corresponds to the circular arc $MCN$ in the bounds given by \cite{milton1981bounds}).}
    \label{fig:Point C locus}
\end{figure}
Following work of Christian Kern (private communication) it is interesting to see what the bounds look in the $y_\epsilon$ plane, where
$y_\epsilon(\epsilon_*) = - f_2 \epsilon_1 
    - f_1 \epsilon_2
    + f_1f_2
    (\epsilon_1 -\epsilon_2 )
    (f_1\epsilon_1  + f_2\epsilon_2 - \epsilon_*)^{-1}
    (\epsilon_1-\epsilon_2)$, is the Y-transform of $\epsilon_*$.
The reason for this is that bounds often simplify when plotted in this plane, and are often volume fraction independent when plotted in this plane. The region $\Psi$ is indeed simpler in the $y_\epsilon$ plane and forms a truncated wedge shaped region bounded by the red straight lines and the line segment $MN$ (Figure \ref{fig:Point C locus}).  
Two sides of this region are attained by the coated ellipsoid geometries, while the point $C$ (after the Y-transformation) traces the third line segment $MN$ as the volume fraction is varied. Thus, the Y-transformation of the conjectured bounds are volume fraction dependent, and thus, cannot be derived using any method that gives volume fraction independent bounds on $y_\epsilon$ (such as the standard analytic method or the standard translation method).

\subsection{Hierarchical Laminates Corresponding to the Conjectured Bounds}
\begin{figure}
    \centering
    \includegraphics[width=\linewidth]{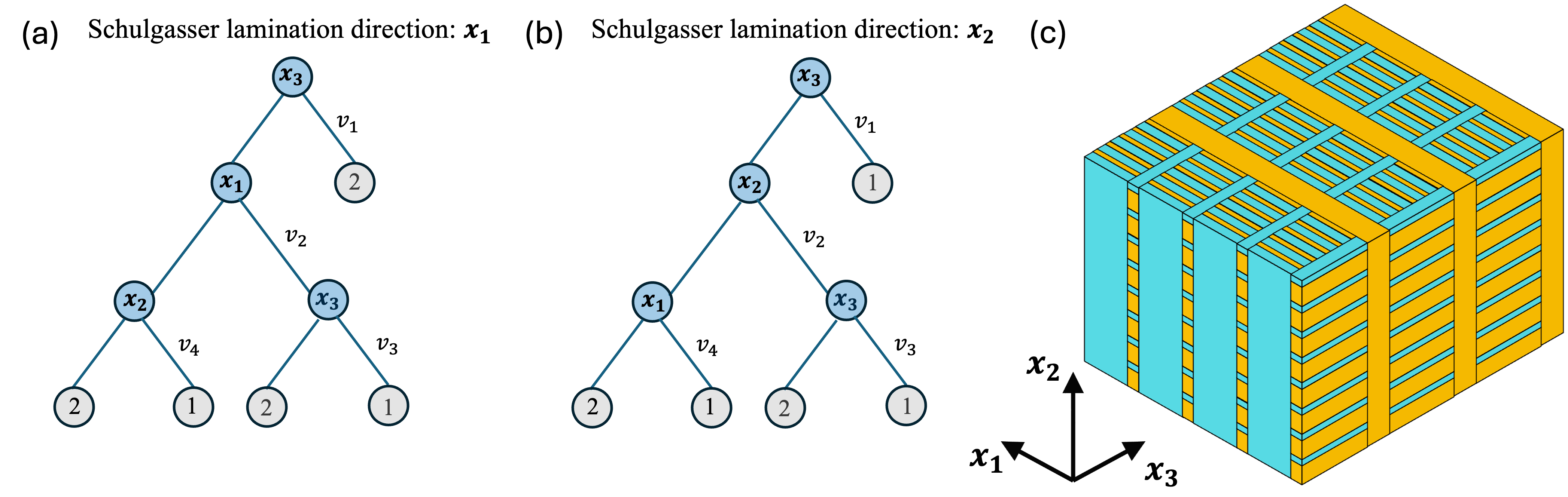}
    \caption{Tree structures of the rank-3 hierarchical laminate (HL) attaining all points on the conjectured bounds when a subsequent Schulgasser lamination step is made.
    (a) Tree structure of rank-3 HL attaining points on the conjectured bound given by blue circular arc $MC$ of $\partial\Gamma$.
    The  numbers inside the grey nodes denote the pure phase, and the blue nodes contain the lamination direction at every step. 
    The general volume fractions of the laminating phases associated with the tree edges are denoted by $v_1, v_2, v_3$, and  $v_4$.
    (b) Tree structure of the rank-3 HL attaining points on the conjectured bound given by red circular arc $NC$ of $\partial\Gamma$.
    (c) The microgeometry of the rank-3 HL laminate represented by the tree structure in (a) is shown with phase-1 in blue and phase-2 in yellow. 
    Schulgasser lamination in the $\bfx_1$ direction with a $90^\circ$ rotation of itself about the $\bfx_1$ axis gives the optimal rank-4 HL.}
    \label{fig:Tree}
\end{figure}
The conjectured bounds are obtained by fitting circles to the $\epsilon_\perp$ values obtained from the Schulgasser laminates of HLs. 
We now focus on finding the specific structure of the hierarchical laminates that attain the points on the conjectured bounds.
A useful way of describing hierarchical laminates is by representing the lamination sequence using a tree structure \citep{milton2002theory,kern2020tight} (see Ch. $9$ in  \cite{milton2002theory}).
Specifically, every hierarchical laminate can be represented as a tree structure in which each node has either zero or two children. The nodes with zero children, commonly known as leaves, correspond to one of the pure phases. 
Conversely, any node that is not a leaf represents a laminate formed from its child nodes and is associated with a particular lamination direction. 
For the three-dimensional HLs considered in this work with subsequent directions of lamination being orthogonal, there are only three possible lamination directions (along the $\bfx_1$, $\bfx_2$, and $\bfx_3$ axes).
In our representation of the tree structure, the numbers inside the leaf nodes correspond to the pure phase (phase-1 or phase-2), and every node that is not a leaf has the lamination direction written in it.
\begin{figure}
    \centering
    \includegraphics[width=\linewidth]{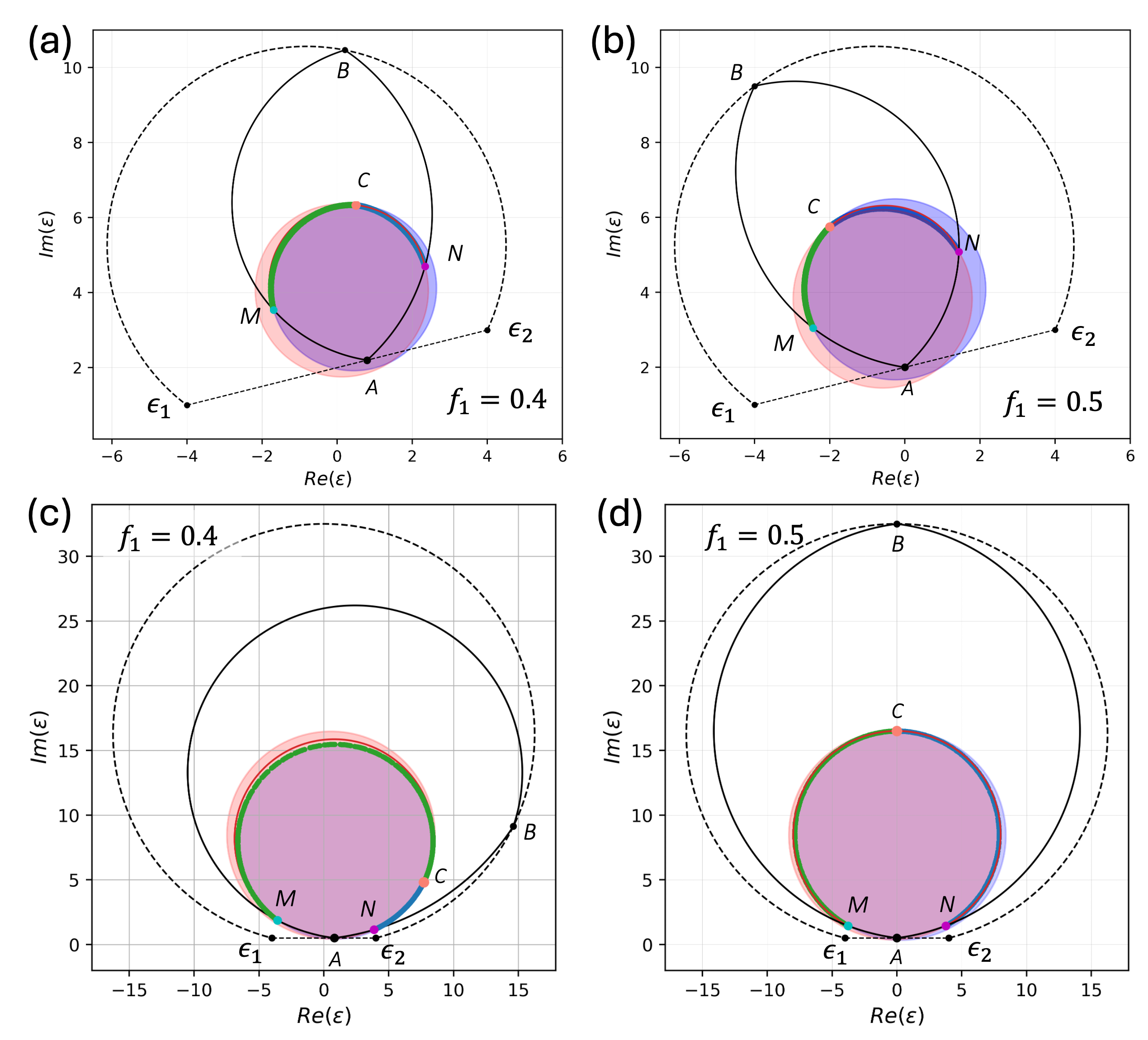}
    \caption{Range of $\epsilon_\perp$ obtained from optimal rank-4 hierarchical laminates attaining green points  on the blue circular arc $MC$ of $\partial\Gamma$ and blue points on the red circular arc $NC$ of $\partial\Gamma$  of the conjectured bounds.
    The classical Bergman-Milton bound on $\epsilon_\perp$ is shown by the solid red curve.
    (a) $\epsilon_1=-4+1\imath$, $\epsilon_2=4+3\imath$, and $f_1=0.4$.
    (b) $\epsilon_1=-4+1\imath$, $\epsilon_2=4+3\imath$, and $f_1=0.5$.
    (c) $\epsilon_1=-4+0.5\imath$, $\epsilon_2=4+0.5\imath$, and $f_1=0.4$.
    (d) $\epsilon_1=-4+0.5\imath$, $\epsilon_2=4+0.5\imath$, and $f_1=0.5$.
    }
    \label{fig:Rank 5 HL}
\end{figure}

The $\epsilon_\perp$ values forming the convex hull and attaining the points on the conjectured bounds in Figures \ref{fig:Psi_fill}\sfred{(d)-(e)} were found to be obtained from rank-4 hierarchical laminates. 
Our goal then is to find a rank-4 tree family that realizes all the points on the conjectured bounds  given by the blue circular  arc $MC$ on $\partial\Gamma$ and the red circular arc $NC$ on $\partial\Gamma$. 
Figure \ref{fig:Tree}\sfred{(a)} shows  a rank-3 (that gives a rank-4 HL after forming a Schulgasser laminate along the $\bfx_1$ direction) HL tree family where by appropriately varying the volume fractions $v_1, v_2, v_3, v_4$ (which are as indicated in Figure \ref{fig:Tree}) of the subsequent lamination steps while keeping $f_1$ fixed gives the blue circular arc $MC$ of $\partial\Gamma$.
The volume fraction $f_1$ for the rank-3 tree family in Figure \ref{fig:Tree}\sfred{(a)} is given as, 
\begin{equation}
    f_1 =  \big(v_4(1-v_2) + v_3 v_2\big)(1-v_1) 
\end{equation}

Similarly, Figure \ref{fig:Tree}\sfred{(b)} shows a rank-3 HL tree family that attains the points on the conjectured bound  given by red circular arc $NC$ of $\partial\Gamma$.
The volume fraction $f_1$ for this rank-3 tree family can be obtained from the relation,
\begin{equation}
    f_1 = \big(v_4(1-v_2) +v_3 v_2 \big)(1-v_1) + v_1 
\end{equation}
Varying the volume fractions $v_1, v_2, v_3, v_4$ while keeping $f_1$ fixed, gives $\epsilon_\perp$ values in the interior as well, along with the points attained on the conjectured bounds.
Interchanging phase-1 and phase-2 in the rank-3 HL tree structure of Figure \ref{fig:Tree}\sfred{(a)}  gives that of Figure \ref{fig:Tree}\sfred{(b)}, and vice-versa.
The optimal HL microstructure corresponding to the rank-3 tree (which becomes rank-4 after the Schulgasser laminate construction)  in Figure \ref{fig:Tree}\sfred{(a)} is shown in Figure \ref{fig:Tree}\sfred{(c)} with the phase-1 shown in blue and the phase-2 shown in yellow.

In Figure \ref{fig:Rank 5 HL}\sfred{(a)}, the green points on the conjectured bound, blue circular arc $MC$ of $\partial\Gamma$, are the $\epsilon_\perp$ values obtained from the family of rank-3 HL shown in Figure \ref{fig:Tree}\sfred{(a)} by varying the volume fractions $v_i$ while keeping $f_1=0.4$ fixed for a composite with $\epsilon_1 = -4+1\imath$, $\epsilon_2=4+3\imath$.
Similarly, the blue points on the conjectured bound, red circular arc $NC$ of $\partial\Gamma$, are the $\epsilon_\perp$ values obtained from the family of rank-3 HL shown in Figure \ref{fig:Tree}\sfred{(b)}. 
For the same composite at a different volume fraction $f_1=0.5$, the optimal $\epsilon_\perp$ values attaining points on the conjectured bounds are shown in Figure \ref{fig:Rank 5 HL}(\sfred{(b)}).
Figures \ref{fig:Rank 5 HL}\sfred{(c)-(d)} show the results for a different composite, with $\epsilon_1 = -4+0.5\imath$, $\epsilon_2=4+0.5\imath$, and $f_1=0.4$ and $f_1=0.5$, respectively. 
In each case, the optimal $\epsilon_\perp$ values are attained using the same tree families shown in Figure \ref{fig:Tree}.
Our numerical simulations indicate that the axial components of the optimal HLs form two tiny loops of blue and green curves near the point $A$, but these could be artifacts of numerical computation.


\section{Correlating the Transverse and Axial Components}\label{sec:correlation}
In applications that require tailored uniaxial permittivity tensors, it is highly desirable to know for a given transverse component what are the correlated axial components, and vice-versa.
In this section, we use HL geometries to explore the correlated axial components of a given fixed value of $\epsilon_\perp$. 
A challenge here is to find HL geometries that exactly achieve the value of a given $\epsilon_\perp$.
Suppose for a two-phase composite with fixed volume fractions, we are interested in finding the range of correlated axial components for a given fixed transverse component denoted by $\epsilon_\perp^{0}$.
We first estimate the feasible  range of $\partial^2\epsilon_\perp/\partial\epsilon_1^2$ that  is compatible with the given transverse component $\epsilon_\perp^0$, and then find the range of the correlated feasible region for the axial component.

For two-phase composites with principal permittivities given by  $\lambda_1(\epsilon_1, \epsilon_2)$, $\lambda_2(\epsilon_1, \epsilon_2)$, and  $\lambda_3(\epsilon_1, \epsilon_2)$,  \cite{bergman1978dielectric} obtained the result
\begin{equation}\label{Bergmann second derivative result}
    \sum_{i=1,2,3} \parderivsec{\lambda_i(\epsilon_1,1)}{\epsilon_1}\Big\vert_{\epsilon_1=1} = -2f_1f_2.
\end{equation}
For uniaxial laminates we have two of the principal permittivities equal to each other, so letting $\epsilon_\parallel=\lambda_1$, and $\epsilon_\perp=\lambda_2=\lambda_3$ we get, 
\begin{equation}\label{first constraint}
     2\parderivsec{\epsilon_\perp}{\epsilon_1}\big\vert_{\epsilon_1=1} = -2f_1f_2 - \parderivsec{\epsilon_\parallel}{\epsilon_1}\Big\vert_{\epsilon_1=1}.
\end{equation}
Another constraint on the the second partial derivatives is given by the bounds derived by  \cite{wiener1912},
\begin{equation}\label{second constraint}
    \parderivsec{\lambda_i(\epsilon_1,1)}{\epsilon_1}\Big\vert_{\epsilon_1=1}\leq 0,
\end{equation}
for $i=1,2,3$.
Using \eqref{first constraint} and \eqref{second constraint} we get the result \citep{milton1981bounds}, 
\begin{equation}\label{trans constraint}
    0 \geq 
    \parderivsec{\epsilon_\perp}{\epsilon_1}\Big\vert_{\epsilon_1=1} 
    \geq 
    -f_1f_2
\end{equation}

Another corollary of \eqref{Bergmann second derivative result} is that isotropic two-phase composites having isotropic permittivity $\epseff$, have the property 
\begin{equation}\label{isotropic constraint}
    \parderivsec{\epseff(\epsilon_1,1)}{\epsilon_1}\Big\vert_{\epsilon_1=1} =-2f_1f_2 / d,
\end{equation}
where the $d$ is the dimension \citep{bergman1978dielectric}. 
\cite{PhysRevLett.45.148.2}
and \cite{milton1980bounds} obtained bounds on the complex isotropic permittivity of a two-phase composite in the isotropic case, i.e., when information about the second partial derivative is known.
These bounds can be written in terms of the second partial derivatives and give the well-known lens shaped bounds.
One of the arcs of the lens is given by the curve traced by $\epsilon_{+}(u_1)$, 
\begin{equation}\label{e5 curve}
    \epsilon_{+}(u_1)
    =
    f_1\epsilon_1
    +f_2\epsilon_2
    -
    \Big(\frac{-1}{2}\parderivsec{\epseff}{\epsilon_1}\Big\vert_{\epsilon_1=1}\Big)
    \frac{(\epsilon_2-\epsilon_1)^2}{(u_1\epsilon_1+u_2\epsilon_2)}
\end{equation}
as $u_1$ is varied so that, $ 1 + \frac{1}{2f_2}\parderivsec{\epseff}{\epsilon_1}\Big\vert_{\epsilon_1=1} \geq u_1 \geq \frac{-1}{2f_2}\parderivsec{\epseff}{\epsilon_1}\Big\vert_{\epsilon_1=1}$, and $u_1+u_2=1$.
The other arc that completes the bounds is given by $\epsilon_{-}(u_1)$, 
\begin{equation}\label{e6 curve}
    \epsilon_{-}(u_1)
    =
    \bigg(
    f_1/\epsilon_1
    +f_2/\epsilon_2
    +
    \Big(-f_1f_2 - \frac{1}{2}\parderivsec{\epseff}{\epsilon_1}\Big\vert_{\epsilon_1=1}\Big)
    \frac{(1/\epsilon_2-1/\epsilon_1)^2}{(u_1/\epsilon_1+u_2/\epsilon_2)}
    \bigg)^{-1},
\end{equation}
where $u_1$ is varied so that, $ f_2 -\frac{1}{2f_2}\parderivsec{\epseff}{\epsilon_1}\Big\vert_{\epsilon_1=1} \geq u_1 \geq f_2 +\frac{1}{2f_1} \parderivsec{\epseff}{\epsilon_1}\Big\vert_{\epsilon_1=1} $, and $u_1+u_2=1$.
\begin{figure}
    \centering
    \includegraphics[width=\linewidth]{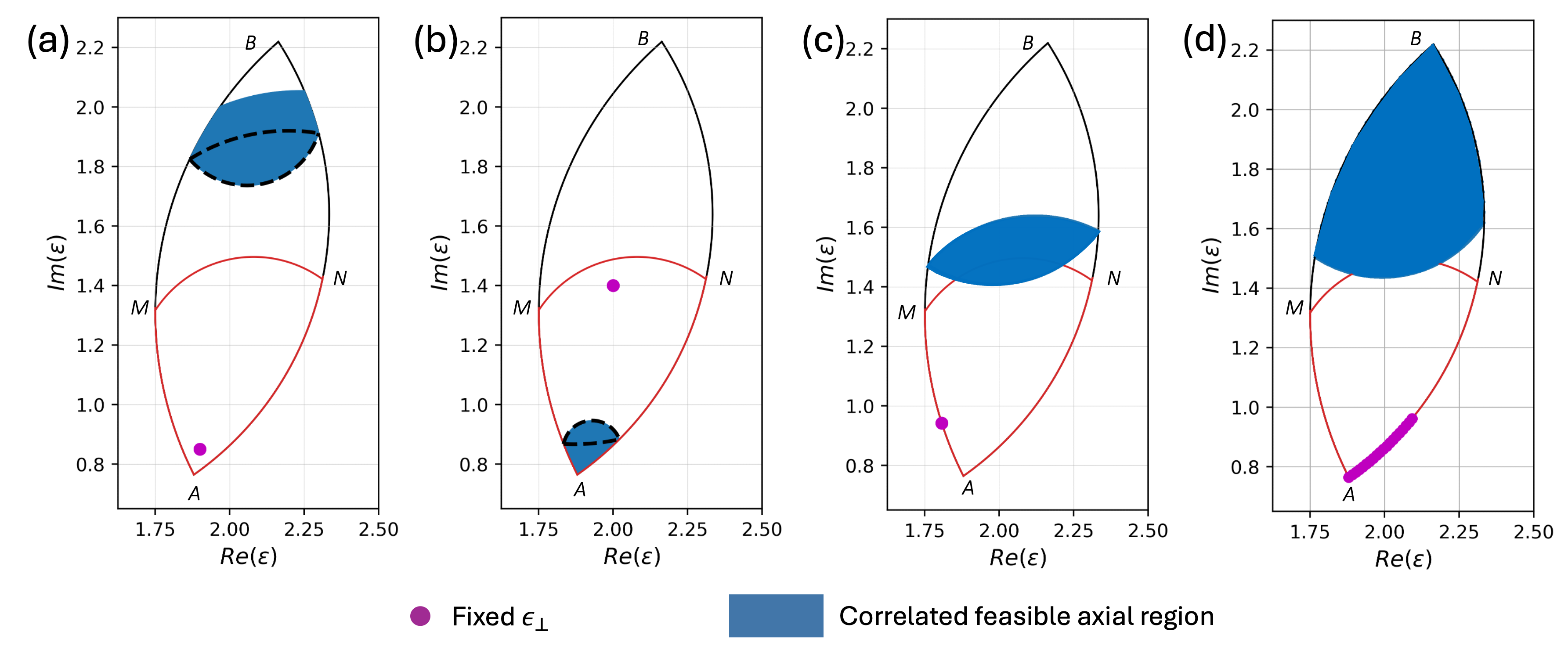}
    \caption{The correlated feasible regions for the axial components of a two phase composite with $\epsilon_1=0.2+1.76\imath$, $\epsilon_2=3+0.1\imath$, and $f_1=0.4$.
    The fixed transverse component $\epsilon_\perp^0$ is shown by the purple marker. 
    The feasible region for the range of correlated axial components is shown in blue.
    The blue regions are formed by the union of the lens shaped regions of the type shown by the dashed black curve.
    (a) For fixed  $\epsilon_\perp^0=1.9+0.85\imath$ in the interior of the region $\Psi$.
    (b) For fixed $\epsilon_\perp^0=2+1.4\imath$ in the interior of the region $\Psi$.
    (c) For fixed $\epsilon_\perp^0$ on the boundary $AM$.
    (d) For $\epsilon_\perp^0$ forming part of the boundary $AN$.}
    \label{fig:feasible}
\end{figure}

\subsection{Axial Components Correlated to a Given Transverse Component}
The main idea now is to use the bounds in an inverse manner to bound the second derivative $\parderivsec{\epseff}{\epsilon_1}$, when the transverse component $\epsilon_\perp^0$ is given.
For every value of $\parderivsec{\epseff}{\epsilon_1}\Big\vert_{\epsilon_1=1}$, \eqref{e5 curve} and \eqref{e6 curve} give two arcs forming a lens shaped region within the region $\Omega^\prime$.
For isotropic composites as stated in \eqref{isotropic constraint} we have a single fixed value of $\parderivsec{\epseff}{\epsilon_1}\Big\vert_{\epsilon_1=1}$ which gives us through relations \eqref{e5 curve} and \eqref{e6 curve} a single lens shaped region, which has been well-described in previous works.
For the transverse and axial components of a uniaxial composite, however, there is a range of possible values that $\parderivsec{\epsilon_\perp}{\epsilon_1}\Big\vert_{\epsilon_1=1}$ and $\parderivsec{\epsilon_\parallel}{\epsilon_1}\Big\vert_{\epsilon_1=1}$ can take as described by the relations \eqref{first constraint} and \eqref{trans constraint}.

For a given fixed transverse component $\epsilon_\perp^0$, we vary $\parderivsec{\epsilon_\perp}{\epsilon_1}\Big\vert_{\epsilon_1=1}$ within the limits given by \eqref{trans constraint} and select only those values of $\parderivsec{\epsilon_\perp}{\epsilon_1}$ such that the associated lens shaped regions contain the point $\epsilon_\perp^0$.
Using \eqref{first constraint}, we can find the $\parderivsec{\epsilon_\parallel}{\epsilon_1}\Big\vert_{\epsilon_1=1}$ value corresponding to each  $\parderivsec{\epsilon_\perp}{\epsilon_1}\Big\vert_{\epsilon_1=1}$ value that is compatible with $\epsilon_\perp^0$.
For these second partial derivatives of the axial component we again plot the lens shaped regions and take their union to get the feasible region for the correlated axial component.

\begin{figure}[b!]
    \centering
    \includegraphics[width=\linewidth]{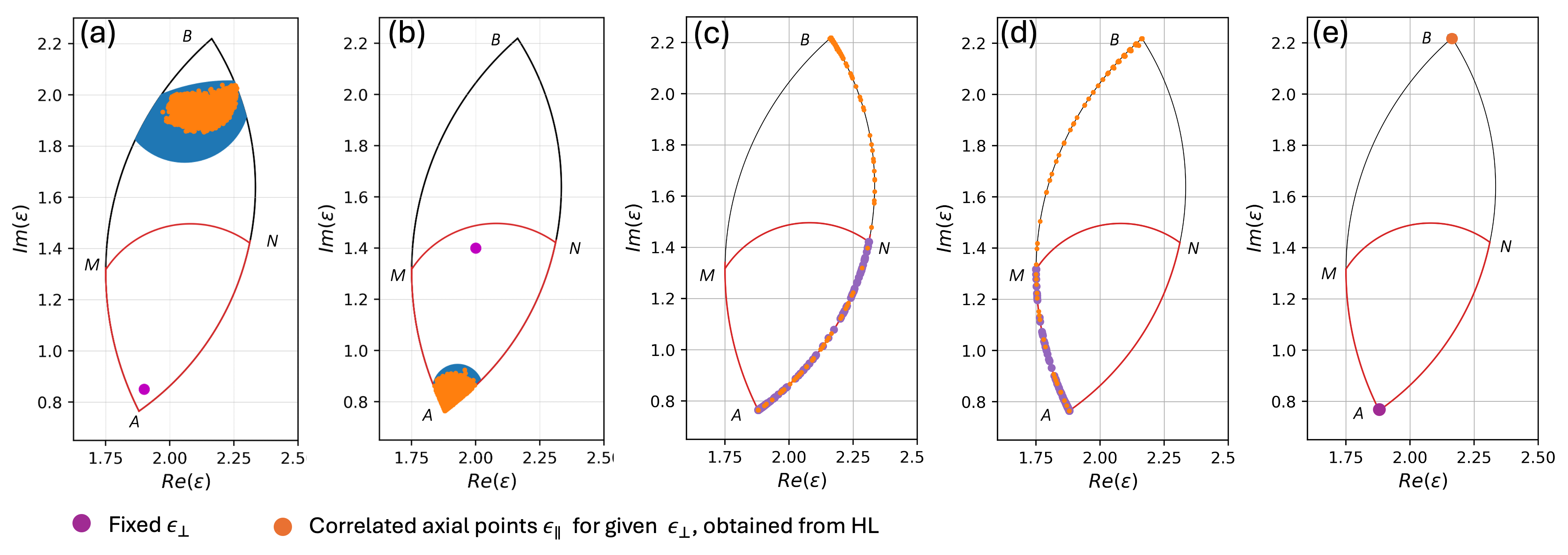}
    \caption{The correlated axial components of the given fixed transverse components obtained from the Schulgasser laminates of HL geometries for a composite with $\epsilon_1=0.2+1.76\imath$, $\epsilon_2=3+0.1\imath$, and $f_1=0.4$.
    (a)-(b)
    The  axial feasible regions are shown in blue color. 
    The fixed transverse component is shown by the purple marker, which is taken to be $\epsilon_\perp^0=1.9+0.85\imath$ for (a), and $\epsilon_\perp^0=2+1.4\imath$ for (b).
    The range of the axial components of the Schulgasser laminates of HL geometries correlated with the fixed transverse component is shown by the orange markers.
    (c)-(d) For fixed $\epsilon_\perp^0$ (purple markers) on the boundaries $AN$ and $AM$, the correlated axial components $\epsilon_\parallel$ (orange markers) lie on the boundary $ANB$ and $AMB$, respectively.
    (e) For $\epsilon_\perp^0=1.9+0.85\imath$ equal to the point $A$, the correlated axial component $\epsilon_\parallel$ is found to be the point $B$.}
    \label{fig:correlated trans to axial}
\end{figure}
For a composite with $\epsilon_1=0.2+1.76\imath$, $\epsilon_2=3+0.1\imath$, and $f_1=0.4$, Figure \ref{fig:feasible}\sfred{(a)} shows the given transverse component $\epsilon_\perp^0$ (shown by the purple marker) in the interior of the region $\Psi$. 
The correlated axial feasible region is shown by the blue region which is formed by the union of the lens shaped regions compatible with the range of values of  $\parderivsec{\epsilon_\parallel}{\epsilon_1}$.
Figure \ref{fig:feasible}\sfred{(b)} shows the correlated feasible regions of the transverse and axial components for a different point in the interior.
In Figure \ref{fig:feasible}\sfred{(c)}, for a point on the boundary $AM$ of the region $\Psi$, we get only one value of $\parderivsec{\epsilon_\perp}{\epsilon_1}$ compatible with the point $\epsilon_\perp^0$ (purple marker) and the correlated axial feasible region is also given by a single lens shaped region (in blue).
The correlated feasible regions for a part of the boundary $AN$ are shown in Figure \ref{fig:feasible}\sfred{(d)}.

For a given fixed transverse component $\epsilon_\perp^0$, our goal now is to find HL geometries that achieve exactly the transverse component $\epsilon_\perp^0$ and then find the corresponding axial components for each of these HL geometries.
We develop a triangulation algorithm to realize HL geometries that give exactly the desired value $\epsilon_\perp^0$ of the transverse component provided that $\epsilon_\perp^0$ lies in the interior of the region $\Psi$.
The main idea of the algorithm is briefly stated in \ref{alg:triangulation_algorithm}.
\begin{algorithm}[h!]
\caption{Triangulation algorithm to generate HL microstructures with a specified value of the transverse component $\epsilon_\perp^0$}
\label{alg:triangulation_algorithm}
\begin{enumerate}
    \item First generate a large number of HL geometries  so that the range of $\epsilon_\perp$ values sufficiently populates the region $\Psi$.
    \item Randomly choose three points from the range of generated $\epsilon_\perp$ points to form a triangle and determine if $\epsilon_\perp^0$ lies inside the triangle.
    \item $\epsilon_\perp^0$ lies inside the triangle if its barycentric coordinates lie within the range $[0,1]$ and if their sum is less than $1$.
    \item If the point is in the interior, then a higher rank HL is constructed by laminating the HLs (which are uniaxial) corresponding to the vertices of the triangle with the lamination direction such that all the three HLs have their axes of symmetry  along the lamination direction.
    \item The volume fractions for the HLs are chosen as the barycentric coordinates of the point $\epsilon_\perp^0$.
\end{enumerate}
\end{algorithm}
This construction method ensures that the new HL formed is also axially symmetric.
However, this scheme will not work if $\epsilon_\perp^0$ lies on the boundary of the region $\Psi$.

Figure \ref{fig:correlated trans to axial}\sfred{(a)} shows the range of correlated axial components (shown by orange markers) for a given transverse component (shown by the purple marker) obtained from HL geometries.
Note that these points lie within the  blue region described in the previous section and illustrated in Figure \ref{fig:feasible}.
Figure \ref{fig:correlated trans to axial}\sfred{(b)} shows the correlated axial components for a different transverse point in the interior of the region $\Psi$.
For $\epsilon_\perp^0$ on the boundary, HL geometries with the exact fixed transverse component may be very difficult to find.
In this case, we generate a large number of HL geometries and  rely on obtaining values of the transverse component within a ball of certain radius centered at $\epsilon_\perp^0$. 
The radius of the ball is suitably chosen to achieve desired numerical accuracy and ideally should tend to $0$.
Figure \ref{fig:correlated trans to axial}\sfred{(c)} shows that for $\epsilon_\perp^0$ points (purple markers) along the boundary $AN$, the correlated axial points (orange markers) are observed to lie along the arc $ANB$.
Similarly, Figure \ref{fig:correlated trans to axial}\sfred{(d)} shows that the correlated axial components of the transverse components on the boundary $MN$, lie on the arc $AMB$.
Finally in Figure \ref{fig:correlated trans to axial}\sfred{(e)}, we observe that the only correlated axial component of the transverse component given by point $A$ is the point $B$.

\begin{figure}
    \centering
    \includegraphics[width=\linewidth]{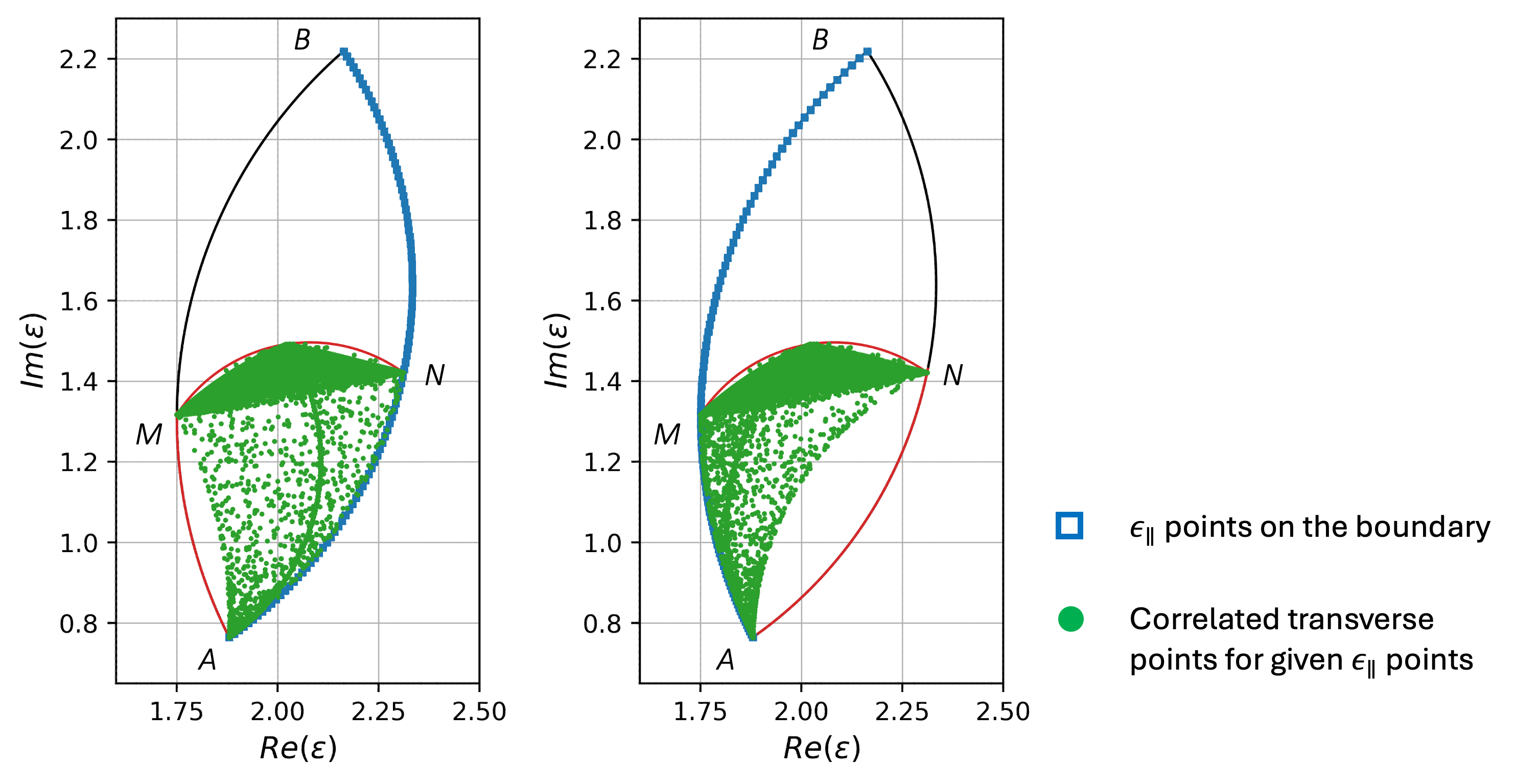}
    \caption{The correlated transverse components of the given fixed axial components obtained from HL geometries for a composite with $\epsilon_1=0.2+1.76\imath$, $\epsilon_2=3+0.1\imath$, and $f_1=0.4$.
    (a)-(b) For fixed $\epsilon_\parallel^0$ (blue markers) on the boundaries $ANB$ and $AMB$, the correlated transverse components (green markers) lie on the boundary $AN$ and boundary $AM$, respectively, as well as in the interior of the region $\Psi$.}
    \label{fig:correlated axial to trans}
\end{figure}

One can perform a similar analysis to obtain correlated transverse components for a given fixed axial component denoted by $\epsilon_\parallel^0$.
Here, we consider just two cases of fixed axial components on the boundaries $AMB$ and $ANB$.
We observe from Figures \ref{fig:correlated axial to trans}\sfred{(a)-(b)} that the correlated transverse components of the given fixed axial components (shown in blue)  on the boundary $ANB$ ($AMB$), lie on the boundary $AN$ ($AM$) as well as in the interior of the region $\Psi$. 


\section{Bounds on the Complex Uniaxial Permittivity Using the Method of Translations}\label{sec:tight}
Our analysis so far suggests that, the bounds of \cite{milton1981bounds} on the complex  uniaxial effective permittivity of two-phase composites are not optimal, specifically the bound given by the boundary $MN$ of the region $\Psi$ shown in Figure \ref{fig:Bounds Milton1981}.
Using the  \cite{cherkaev1994variational} transformation and the method of translations \citep{tartar2018estimations,lurie1984exact,murat_tartar_1985}, \cite{kern2020tight}   derived an optimal bound on the complex isotropic effective permittivity of two-phase composites.
Following a similar procedure we derive bounds on the complex effective  permittivity of a uniaxial two-phase composite. 
The constitutive relations relating the electric field $\bfe(\bfx)$ and the electric displacement $\bfd(\bfx)$ are rewritten in terms of the real parts (primed) and the imaginary parts (double primed) of the field quantities as, 
\begin{equation}
    \begin{bmatrix}
        \bfe{''} 
        \\
        \bfd{''}
    \end{bmatrix}
    =
    \begin{bmatrix}
        (\bfepsilon{''})^{-1} & (\bfepsilon{''})^{-1}\bfepsilon{'} 
        \\
        \bfepsilon{'} (\bfepsilon{''})^{-1} & \bfepsilon{'} (\bfepsilon{''})^{-1} \bfepsilon{'}  + \bfepsilon{''}
    \end{bmatrix}
    \begin{bmatrix}
        -\bfd{'} 
        \\
        \bfe{'}
    \end{bmatrix}
    = \bfL 
    \begin{bmatrix}
        -\bfd{'} 
        \\
        \bfe{'}
    \end{bmatrix},
\end{equation}
where $\bfL$ is a symmetric and positive definite second-order tensor. 
The corresponding minimization principle \citep{cherkaev1994variational} to find the effective permittivity tensor can be stated as
\begin{equation}\label{variational problem rank-2}
\begin{bmatrix}
    - \bfd_0' \\
\bfe_0'
\end{bmatrix}
\cdot \bfL_* 
\begin{bmatrix}
    - \bfd_0' \\
\bfe_0'
\end{bmatrix}
= 
\min 
\left\{ \left\langle 
\begin{bmatrix}
- \bfd' \\
\bfe'    
\end{bmatrix}
\cdot \bfL 
\begin{bmatrix}
- \bfd' \\
\bfe'    
\end{bmatrix}
\right\rangle \ \middle| \
\begin{array}{l}
\langle \bfd' \rangle = \bfd_0', \quad \nabla \cdot \bfd' = 0 \\
\langle \bfe' \rangle = \bfe_0', \quad \nabla \times \bfe' = 0
\end{array} \right\},
\end{equation}
where $\langle\cdot\rangle$ denotes a volume average.
Following the arguments and methods used in  \cite{kern2020tight}, the problem in \eqref{variational problem rank-2} is embedded in a problem involving a fourth-order tensor $\bfcalL$ comprising several copies of $\bfL$.
Let $\mathbf{\cal{E}}$ denote the fourth-order tensor given in the component form by
\begin{equation}\label{rank-4 calE}
    \calE_{ijkl} = \epsilon_{ik}\delta_{jl}, \quad i,j,k,l=1,2,3,
\end{equation}
then the tensor $\bfcalL$ is given in terms of $\calE$ as,
\begin{equation}
    \bfcalL = 
    \begin{bmatrix}
        (\mathbf{\cal{E}}{''})^{-1} & (\mathbf{\cal{E}}{''})^{-1}\mathbf{\cal{E}}{'} 
        \\
        \mathbf{\cal{E}}{'} (\mathbf{\cal{E}}{''})^{-1} & \mathbf{\cal{E}}{'} (\mathbf{\cal{E}}{''})^{-1} \mathbf{\cal{E}}{'}  + \mathbf{\cal{E}}{''}
    \end{bmatrix}.
\end{equation}
It is convenient to introduce the Y-transform (see Ch. 19 in  \cite{milton2002theory} and references therein) of the effective tensor $\mathbf{\calE}_*$ given by the relation, 
\begin{equation}\label{y transform of calE}
    \mathbf{\calY} = -f_1\mathbf{\calE}_2 -f_2\mathbf{\calE}_1
    +f_1f_2 
    (\mathbf{\calE}_1-\mathbf{\calE}_2)
    \cdot
    (f_1\mathbf{\calE}_1 +f_2\mathbf{\calE}_2 - \mathbf{\calE}_*)^{-1}
    (\mathbf{\calE}_1-\mathbf{\calE}_2),
\end{equation}
using which we can write the Y-transform of the effective tensor $\bfcalL_*$ as, 
\begin{equation}\label{y transform of L}
    \bfY = 
    \begin{bmatrix}
        (\mathbf{\calY}{''})^{-1} & -(\mathbf{\calY}{''})^{-1}\mathbf{\calY}{'} 
        \\
        -\mathbf{\calY}{'} (\mathbf{\calY}{''})^{-1} & \mathbf{\calY}{'} (\mathbf{\calY}{''})^{-1} \mathbf{\calY}{'}  + \mathbf{\calY}{''}
        \end{bmatrix}.
\end{equation}
We omit a detailed discussion of the procedure, which is discussed in sufficient detail in  \cite{kern2020tight}, and instead directly use the important results. 
Using the Y-transform $\bfY$ we get the simple bounds, 
\begin{equation}\label{translation bounds}
    \bfY + \bfT \geq 0,
\end{equation}
where the translation tensor $\bfT$ is chosen such that $\bfT$ is quasiconvex \citep{murat_tartar_1985,lurie1984exact,Tartar1985,gibiansky1993effective} and the translated tensors $\bfcalL_1-\bfT \geq0$ and $\bfcalL_2-\bfT \geq0$ are positive semi definite.
We choose the same form for $\bfT$ as the one used in  \cite{kern2020tight} that is 
\begin{equation}
    \bfT = 
    \begin{bmatrix}
        \bfA(-t_1,2t_1,0) & \bfA(-t_3,-t_3,-t_3)
        \\
      \bfA(-t_3,-t_3,-t_3) & \bfA(-2t_2,t_2,-t_2),
    \end{bmatrix}
\end{equation}
where $t_1\geq 0 $, $t_2,t_3$ are arbitrary, and $\bfA$ is a general fourth-order isotropic tensor given in the component form by 
\begin{equation}
A_{ijkl}(t_1, t_2, t_3) = \frac{t_1}{3} \delta_{ij} \delta_{kl} 
+ \frac{t_2}{2} \left( \delta_{ik} \delta_{jl} + \delta_{il} \delta_{jk} - \frac{2}{3} \delta_{ij} \delta_{kl} \right)
+ \frac{t_3}{2} \left( \delta_{ik} \delta_{jl} - \delta_{il} \delta_{jk} \right).
\end{equation}

The effective uniaxial permittivity $\bfepseff$ can be decomposed as linear combination of two orthogonal tensors
\begin{equation}
    \bfepseff = \epsilon_\perp (\bfI - \bfn \otimes\bfn) + \epsilon_\parallel \bfn \otimes\bfn,
\end{equation}
using which we can write the fourth-order tensor $\mathbf{\calE}_*$ from \eqref{rank-4 calE} in the component form as
\begin{equation}
    {\calE_*}_{ijkl}
    =
    \epsilon_\perp (\delta_{ik} - n_in_k)\delta_{jl}
    + \epsilon_\parallel n_in_k\delta_{jl}.
\end{equation}
Here, $\bfn$ is the direction parallel to the axis of symmetry of the uniaxial composite.
Substituting the above in \eqref{y transform of calE} the Y-transform of the above fourth-order tensor is given by
\begin{equation}\label{y transform of calE in decomposed form}
    \calY_{ijkl} = \Lambda_1 A_{ijkl}(1,1,1) + \Lambda_2 n_in_k\delta_{jl} = \Lambda_1(\delta_{ik}-n_in_k)\delta_{jl} + (\Lambda_2+\Lambda_1)n_in_k\delta_{jl},
\end{equation}
where, 
\begin{equation}
\begin{split}
    &\Lambda_1 = -f_1\epsilon_2-f_2\epsilon_1+\frac{f_1f_2(\epsilon_1-\epsilon_2)^2}{f_1\epsilon_1+f_2\epsilon_2-\epsilon_\perp},
    \\
    &\Lambda_2 = -f_1f_2(\epsilon_1-\epsilon_2)^2
    \Big(
    \frac{1}{f_1\epsilon_1+f_2\epsilon_2-\epsilon_\perp}
    -\frac{1}{f_1\epsilon_1+f_2\epsilon_2-\epsilon_\parallel}
    \Big).
    \end{split}
\end{equation}
The second equality in \eqref{y transform of calE in decomposed form} is convenient for getting the inverse of $\mathbf{\calY}$.
It also implies that $\Lambda_1$ is the Y-transformed  transverse component and $(\Lambda_1+\Lambda_2)$ is the Y-transformed axial component.

Next, substitute \eqref{y transform of calE in decomposed form} in \eqref{y transform of L}, and together with the chosen form of the translation $\bfT$, substitute the result in the translation bounds given in \eqref{translation bounds}.
The  positive semi definiteness constraint $\bfY+\bfT \geq 0$ can be simplified to positive semi definiteness conditions on smaller submatrices by identifying the coupling terms. 
These conditions are:
\begin{equation}\label{matrix 1}
    \begin{bmatrix}
        t_1+\frac{1}{(\Lambda_1+\Lambda_2)''} & -\sqrt{2}t_1 & -t_3 - \frac{(\Lambda_1+\Lambda_2)'}{(\Lambda_1+\Lambda_2)''} & 0
        \\
        -\sqrt{2}t_1 & \frac{1}{\Lambda_1''}&0 & -t_3 - \frac{\Lambda_1'}{\Lambda_1''}
        \\
        -t_3 - \frac{(\Lambda_1+\Lambda_2)'}{(\Lambda_1+\Lambda_2)''} & 0& (\Lambda_1+\Lambda_2)''+ \frac{(\Lambda_1+\Lambda_2)'^2}{(\Lambda_1+\Lambda_2)''}& -\sqrt{2} t_2
        \\
        0& -t_3 - \frac{\Lambda_1'}{\Lambda_1''}& -\sqrt{2}t_2& -t_2+ \Lambda_1''+\frac{\Lambda_1'^2}{\Lambda_1''}
    \end{bmatrix}
    \geq
    0,
\end{equation}
\begin{equation}\label{matrxi 2}
    \begin{bmatrix}
        2 t_1 +\frac{1}{\Lambda_1''} & -t_3 - \frac{\Lambda_1'}{\Lambda_1''}
        \\
        -t_3-\frac{\Lambda_1'}{\Lambda_1''}& t_2+ \Lambda_1''+\frac{\Lambda_1'^2}{\Lambda_1''}
    \end{bmatrix}
    \geq
    0,
\end{equation}
and 
\begin{equation}\label{matrix 3}
    \begin{bmatrix}
        \frac{1}{\Lambda_1''}&0& -t_3-\frac{\Lambda_1'}{\Lambda_1''}&0
        \\
        0& \frac{1}{\Lambda_1''} & 0& -t_3-\frac{\Lambda_1'}{\Lambda_1''}
        \\
        -t_3-\frac{\Lambda_1'}{\Lambda_1''} &0& \frac{|\Lambda_1|^2}{\Lambda_1''}&-t_3
        \\
        0&-t_3-\frac{\Lambda_1'}{\Lambda_1''} &-t_3&\frac{|\Lambda_1|^2}{\Lambda_1''}
    \end{bmatrix}
    \geq
    0.
\end{equation}
These conditions cannot be further simplified, and they imply that the determinants of the two $4\times4 $ matrices and the one $2\times2$ matrix are non-negative.
The translation parameters $t_1, t_2, t_3$ are chosen so that the translated tensors $\bfcalL_i-\bfT$ are positive semi definite, which implies the following two necessary and sufficient non-negativity conditions on the translation parameters \citep{kern2020tight}
\begin{equation}\label{condition 1 on translation parameter}
    (y_c^\prime + \epsilon_i^\prime)^2 + (y_c^\dprime + \epsilon_i^\dprime)^2 
    \geq R^2,
\end{equation}
and 
\begin{equation}\label{condition 2 on translation parameter}
    (y_c^\prime - 2\epsilon_i^\prime)^2 + (y_c^\dprime - 2\epsilon_i^\dprime)^2 
    \leq R^2,
\end{equation}
where
\begin{equation}
    y_c^\prime = -\frac{t_3}{t_1}, 
    \quad 
    y_c^\dprime = -\frac{1+2t_1t_2-t_3^2}{2t_1}, 
    \quad
\text{and }    R = \left\vert\frac{1-2t_1t_2+t_3^2}{2t_1}\right\vert.
\end{equation}
These restrictions correspond to choice of $t_1, t_2, t_3$ such that $2\epsilon_1$ and $2\epsilon_2$ do not lie outside of the circle and $-\epsilon_1$ and $-\epsilon_2$ do not lie inside of the circle.
The extremal translation parameters consistent with the restrictions correspond to the halfspace bounded by a straight line between $2\epsilon_1$ and $2\epsilon_2$ that does not contain $-\epsilon_1$ and $-\epsilon_2$, and one of the two circles passing through $2\epsilon_1, 2\epsilon_2$, and $-\epsilon_1$ or $-\epsilon_2$.
This extremal choice of translations  was obtained in  \cite{kern2020tight}, and is directly used for our case since we are using exactly the same translation tensor.

We observe through numerically testing the constraints \eqref{matrix 1}-\eqref{matrix 3}, that the particular choice of the isotropic translation tensor considered in this work does not give tighter bounds on $\epsilon_\perp$ when $\epsilon_\parallel$ in unknown or on $\epsilon_\parallel$ when $\epsilon_\perp$ is unknown, and that in fact these bounds are poorer than the original bounds given by \cite{milton1981bounds}.
However, they do give bounds correlating $\epsilon_\perp$ and $\epsilon_\parallel$.
The fact that our translation bound does not improve the bounds on $\epsilon_\perp$ and $\epsilon_\parallel$ is to be expected. 
Clearly the bounds on $\epsilon_\perp$ are optimal in the Y-transformed plane. The improved bound in \cite{kern2020tight} in the isotropic case is on the side nearest $A$, so similarly one expects in this case it would be if $\parderivsec{\epsilon_\perp}{\epsilon_1}\Big\vert_{\epsilon_1=1}=-f_1f_2$.

\begin{figure}
    \centering
    \includegraphics[width=0.6\linewidth]{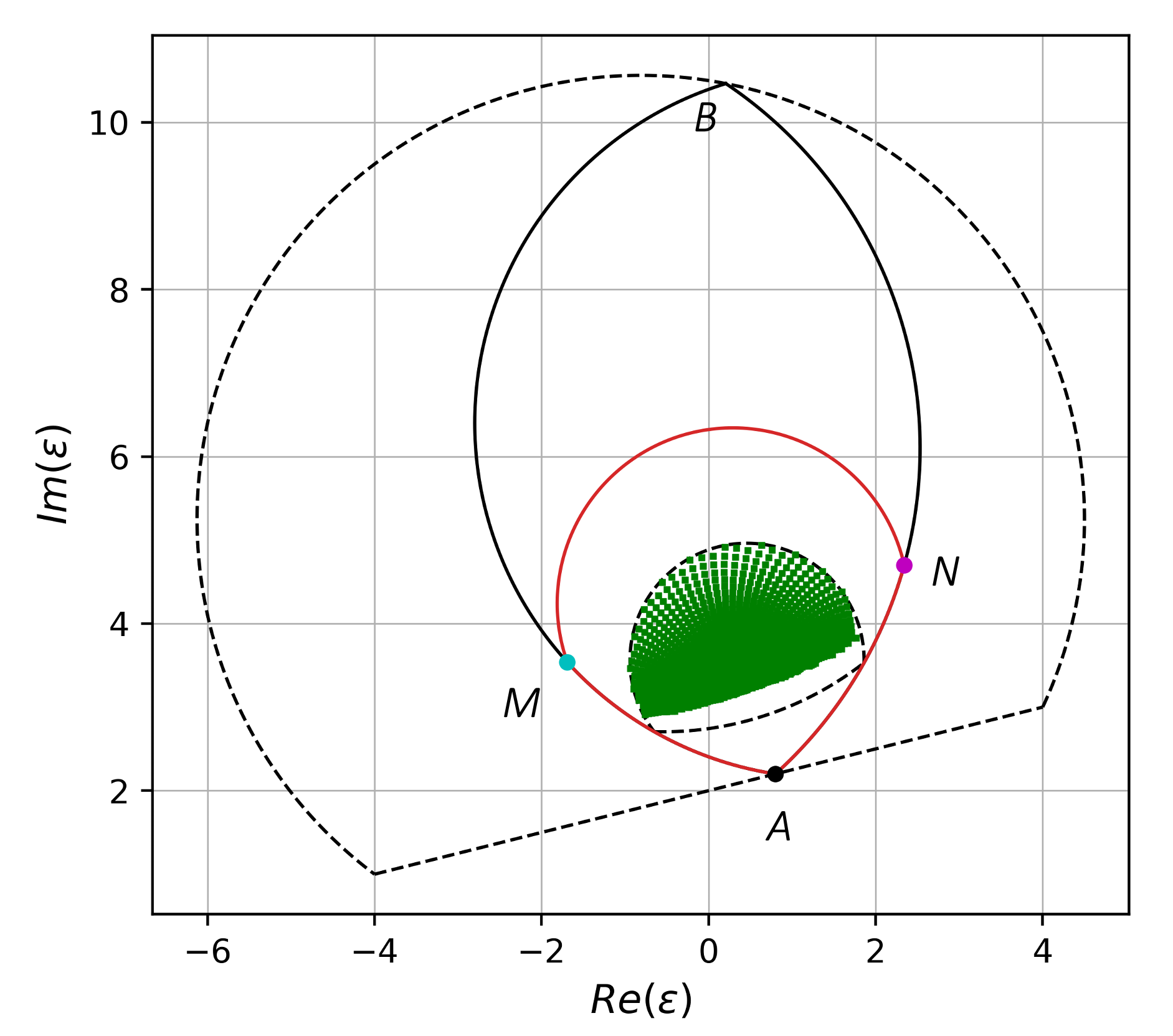}
    \caption{Numerically evaluated translation bounds on $\epsilon_\perp$ shown by the area shaded in green when $\epsilon_\perp$ and $\epsilon_\parallel$ are correlated as $\epsilon_\parallel=1.1\epsilon_\perp$ for a composite with $\epsilon_1=-4+1\imath$, $\epsilon_2=4+3\imath$, and $f_1=0.4$.
    When $\epsilon_\perp = \epsilon_\parallel$, it corresponds to the case of isotropic effective permittivity, and we recover the translation bounds obtained by \cite{kern2020tight} and are shown by the dashed black curves.
    The green region lies within the region enclosed by the  dashed black curves.
    (b) Y-transform of the region $\Psi$ (bounded by the red straight lines and the line passing through points $M$ and $N$) is shown with points $M, N, B$, and $C$ shown in cyan, pink, black, and orange, respectively, for composite with $\epsilon_1=0.2+1.76\imath$, $\epsilon_2=3+0.1\imath$ and unknown volume fraction. 
    In the Y-plane, the region $\Psi$ is volume fraction independent, and the point $C$ attained by the L geometry traces the bound given by the straight line joining points $M$ and $N$ (which corresponds to the circular arc $MCN$ in the bounds given by \cite{milton1981bounds}).
    This shows that $\Psi$ is optimal when the volume fractions are not fixed.}
    \label{fig:tight}
\end{figure}

In the limit $\epsilon_\perp\to\epsilon_\parallel$, we have $\Lambda_2\to0$, and when $\epsilon_\perp=\epsilon_\parallel$ we get the isotropic case. 
The bounds given by \eqref{matrix 1}-\eqref{matrix 3} in this case reduce to the tight  bounds on the effective isotropic permittivity given by \cite{kern2020tight}.
Now assume that the uniaxial permittivity tensor is a perturbation of the isotropic permittivity tensor such that $\epsilon_\parallel = \alpha\epsilon_\perp$, i.e, the transverse component differs from the axial component by a small amount when the scalar $\alpha$ is close to $1$.
For the case when $\alpha=1.1$, we numerically evaluate the translation bounds on $\epsilon_\perp$, which we expect to be a perturbation of the isotropic tight bounds given by \cite{kern2020tight}.
The numerical bounds on $\epsilon_\perp$ obtained from the constraints \eqref{matrix 1}-\eqref{matrix 3}, with the assumption that the transverse and axial components have the correlation $\epsilon_\parallel = 1.1\epsilon_\perp$, are shown in Figure \ref{fig:tight} by the green region.
The lens shaped region given by the dashed black curve are the tight isotropic bounds (when $\epsilon_\perp= \epsilon_\parallel$ or $\alpha=1$) given by \cite{kern2020tight}.
As $\alpha\to1$ the green region extends to fill up the region bounded by the black dashed curves.
Thus, the translation bounds given here in \eqref{matrix 1}-\eqref{matrix 3} are useful at least when $\epsilon_\perp$ and $\epsilon_\parallel$ differ only by a small fixed amount.

\section{Bounds on the Sensitivity of the Complex Effective Permittivity}\label{app:A}
In this section, our goal is to find bounds on the sensitivity of the anisotropic effective permittivity with respect to the phase permittivity in low-loss composites.
We will show that in two-dimensions, bounds on the sensitivity lead to optimal bounds on the effective permittivity in the low-loss limit.
\cite{schulgasser1976bounds} obtained bounds on the effective permittivity of low-loss composites by using bounds on $\partial\bfepseff / \partial \epsilon_i$ ($i=1,2$)  given by \cite{prager1969improved} for the case of statistically homogeneous isotropic two-phase composites with no loss.

\subsection{Known Bounds}
Let the Y-transform $\bfY(\bfepseff)$ of the effective permittivity be given by  
\begin{equation}\label{Y-transform definition}
\begin{split}
    \bfY (\bfepseff) 
    = 
    - f_2 \epsOne 
    - f_1 \epsTwo
    + f_1f_2
    (\epsOne-\epsTwo)
    (f_1\epsOne + f_2\epsTwo - \bfepseff)^{-1}
    (\epsOne-\epsTwo)
\end{split}
\end{equation}
From the analytic properties of the function $\bfepseff(\epsilon_1, \epsilon_2)$ it follows that $\bfY_*(\epsilon_1,\epsilon_2)$ satisfies
\begin{enumerate}
    \item The Herglotz property 
        \begin{equation}
          \IM{[\bfY_*(\epsilon_1, \epsilon_2)]} \geq 0,
        \end{equation}
        when both $\IM{(\epsilon_1)}>0$, $\IM{(\epsilon_2)}>0$.
    \item The homogeneity property
    \begin{equation}
          \bfY_*(\lambda\epsilon_1, \lambda\epsilon_2) 
          =
          \lambda\bfY_*(\epsilon_1, \epsilon_2).
        \end{equation}
\end{enumerate}
Without loss of generality, by relabeling the phases if necessary, we may assume $\IM{[\epsilon_1/\epsilon_2]}>0$.
Letting $\lambda$ appropriately approach $1/\epsilon_2$ we get
\begin{equation}\label{lambda limit 1}
    \IM{[\bfY_*(\epsilon_1/\epsilon_2,1)]}
    =
    \IM{[\bfY_*(\epsilon_1,/\epsilon_2)/\epsilon_2]}
    \geq
    0.
\end{equation}
Similarly, letting $\lambda$ appropriately approach $-1/\epsilon_1$ we get
\begin{equation}\label{lambda limit 2}
    \IM{[\bfY_*(\epsilon_1,/\epsilon_2)/\epsilon_1]}
    \leq
    0.
\end{equation}

Equations \eqref{lambda limit 1} and \eqref{lambda limit 2} are the wedge bounds written in an equivalent form in \cite{milton1987multicomponent} formulae $(4.1)$ to $(4.6)$, where $\Lambda^{(1)}$ should be identified with $\bfY_*$.
These  bounds on $\bfY_*$ naturally imply bounds on $\bfepseff$.

Other bounds  on complex $\bfepseff$ have been derived (\cite{milton1990characterizing} - see also \cite{milton2002theory}, Eqn (22.4)) using the Cherkaev-Gibiansky variantional principle \citep{cherkaev1994variational}.
These bounds with $e^{\imath \theta} = \epsilon_2/|\epsilon_2|$ and $\epsilon_0$ being real and approaching $|\epsilon_2|$, give after dividing them by $|\epsilon_2|$,
\begin{equation}
\begin{split}
    \left\{
    \IM{[(\bfepseff/\epsilon_2-\bfI)^{-1}]}
    \right\}^{-1}
    &\geq
    f_1 
    \left\{
    \IM{[(\epsilon_1/\epsilon_2-1)^{-1}]}
    \right\}^{-1} \bfI +
    f_2
    \left\{
    \IM{[(1 - \epsilon_0/|\epsilon_2|)^{-1}]}
    \right\}^{-1}\bfI
    \\
    &\geq 
    f_1 
    \left\{
    \IM{[(\epsilon_1/\epsilon_2-1)^{-1}]}
    \right\}^{-1}\bfI,
\end{split}
\end{equation}
or equivalently
\begin{equation}\label{C}
    \IM{[(\bfepseff/\epsilon_2-\bfI)^{-1}]}
    \leq
    \IM{[(\epsilon_1/\epsilon_2-1)^{-1}]}\bfI /f_1 .
\end{equation}
Similarly, taking $e^{\imath \theta} = -\epsilon_1/|\epsilon_2|$ and $\epsilon_0$ approaching $-|\epsilon_2|$ we get,
\begin{equation}\label{D}
    \IM{[(\bfepseff/\epsilon_1-\bfI)^{-1}]}
    \geq
    \IM{[(\epsilon_2/\epsilon_1-1)^{-1}]}\bfI / f_2 .
\end{equation}
Now, let us show the equivalence of these bounds with the wedge bounds. 
Using the relation \eqref{Y-transform definition} for $\bfY_*$ we get
\begin{equation}
    \bfepseff - \bfepsilon_2 
    =
    f_1(\bfepsilon_1-\bfepsilon_2)
    [f_2 \bfepsilon_1-f_1\bfepsilon_2+\bfY_*]^{-1}
    [\bfepsilon_2+\bfY_*]
\end{equation}
After inverting and multiplying by $\epsilon_2$ we get
\begin{equation}
    (\bfepseff/\epsilon_2-\bfI)^{-1}
    =
    f_1^{-1}f_2 (\bfI+\bfY_*/\epsilon_2)^{-1}
    +
    f_1^{-1} (\epsilon_1/\epsilon_2-1)\bfI.
\end{equation}
Taking the imaginary part of both sides and using \eqref{C} we get
\begin{equation}
    \IM{[1+\bfY_*/\epsilon_2]^{-1}}
    \leq
    0,
\end{equation}
which implies
\begin{equation}
    \IM{[\bfY_*(\epsilon_1,\epsilon_2)/\epsilon_2]}
    \geq
    0,
\end{equation}
which is the wedge bound given by \eqref{lambda limit 1}.
Similarly, the other wedge bound given by \eqref{lambda limit 2} follows from \eqref{D}.

\subsection{Optimality of the Wedge Bounds for Two-dimensional Low-loss Composites} 
Consider a quasistatic two-phase composite with an anisotropic complex effective  permittivity tensor denoted by $\bfepseff$. 
The constituent phases are assumed to have low loss and the permittivity of phase-1 and phase-2 are denoted by $\epsOne=\epsOneReal+\imath \epsOneIm$ and $\epsTwo=\epsTwoReal+\imath \epsTwoIm$, respectively, where the primes denote the real part and double primes denote the imaginary part. 
Let $f_1$ and $f_2$ denote the volume fractions of phase-1 and phase-2 respectively.
The low loss assumption implies that, $\epsOneIm<<\epsOneReal, \epsTwoIm<<\epsTwoReal$.

Without loss of generality (due to homogeneity) we assume the second phase has a constant real-valued dielectric permittivity with zero loss, $\epsTwo=\epsTwoReal>0$.
Since, $\imath \epsOneIm$ can also be viewed as a small imaginary perturbation to the permittivity $\epsOne=\epsOneReal>0$, we first derive our bounds on the sensitivity when the perturbation is real, and later justify the extension to the case when the perturbation is imaginary (the low loss limit) which as we will show is very straightforward.
Hence, the permittivity of the first phase (along with the second phase) is now taken to be real-valued with $\epsOneIm$ being a small perturbation to the permittivity $\epsOne$, i.e., $\epsOne=\epsOneReal + \epsOneIm$.
Our main goal is to now bound the sensitivity $\partial\bfepseff/\partial\epsilon_1$.

We formulate our problem in two-dimensions as one of finding the coupled bounds on the effective tensors of two properties of a two-phase composite by using the Cherkaev-Gibiansky transformation  \citep{cherkaev1992exact}.
Exact coupled bounds for effective tensors of electrical and magnetic properties in a 2d two-phase composite were given by \cite{cherkaev1992exact}.
We follow the same procedure and use the subsequent results derived therein to find bounds on the sensitivity.

Let $\epsOne=\epsilon_1^\prime \bfI$ and $\epsTwo=\epsilon_2^\prime \bfI$ denote the isotropic unperturbed dielectric permittivity tensors.
Now let $\bfmu_1 = \epsOneReal+\epsOneIm$ denote the perturbed dielectric permittivity of phase-1  for a small real perturbation  ($\epsOneIm$).
For simplicity, we define $\bfmu_2 = \epsTwoReal$ to be the isotropic dielectric permittivity of phase-2 which remains unchanged, i.e., $\bfmu_2=\epsTwo=\epsTwoReal$.

The effective tensors of the unperturbed and perturbed
anisotropic dielectric permitivitties are denoted by $\epseff(\epsOne,\epsTwo)$ and $\bfmu_*(\bfmu_1, \bfmu_2)$, respectively. 
\cite{cherkaev1992exact} gave relations that will be satisfied by all possible pairs of the effective tensors $\bfepseff$ and $\bfmu_*$. 
In particular, when the phase properties satisfy the inequality, 
\begin{equation}\label{phase case B}
    (\epsilon_1-\epsilon_2)(\mu_1-\mu_2)
    \geq 
    0,
\end{equation}
then the effective tensors satisfy the relation, 
\begin{equation} \label{CG relation 2}
    \frac{\epsilon_1}{\mu_1} \bfY(\bfmu_*)
    \leq
    \bfY(\bfepseff)
    \leq 
    \frac{\epsilon_2}{\mu_2} \bfY(\bfmu_*),
\end{equation}
provided that $\frac{\epsilon_1}{\mu_1}\leq\frac{\epsilon_2}{\mu_2}$.
Note that the relation as stated in their original paper
\begin{equation}\label{Cherkaev_Gibiansly relation 1}
    \epsOne \bfmu_1^{-1} \leq \bfY(\bfepseff)\bfY(\bfmu_*)^{-1}
    \leq 
    \epsTwo\bfmu_2^{-1},
\end{equation}
is incorrect, because it assumes that the tensors $\bfY(\bfepseff)$ and $\bfY(\bfmu_*)$ commute which is not true in general.
The correct form of the relation is given in \eqref{CG relation 2}.

Before proceeding further, we verify that the small perturbations considered here correspond to the case given by the inequality in \eqref{phase case B}. 
With $\epsilon_1 = \epsilon_1^\prime$, $\mu_1 = \epsilon_1^\prime +\epsilon^\dprime$, and  $\mu_2 = \epsilon_2 = \epsilon_2^\prime$ substituted in \eqref{phase case B} we have, 
\begin{equation}\label{phase case B 2}
\begin{split}
        &(\epsilon_1^\prime-\epsilon_2^\prime)(\epsilon_1^\prime+\epsilon_1^\dprime-\epsilon_2^\prime)
        \geq 
        0,
        \\
        \implies
        &
        (\epsilon_1^\prime-\epsilon_2^\prime)^2 
        +
        (\epsilon_1^\prime-\epsilon_2^\prime) \epsilon_1^\dprime
        \geq 
        0.
    \end{split}
\end{equation}
We can see that for arbitrary small perturbations $\epsilon_1^\dprime<<\epsilon_1^\prime$, the inequality in \eqref{phase case B 2} holds. 
Contrast this with the Cherkaev-Gibiansky bounds  given in the case when the component phases satisfy the relation $(\epsilon_1-\epsilon_2)(\mu_1-\mu_2)\leq 0$. 
For the problem in this section, this translates to $(\epsilon_1^\prime-\epsilon_2^\prime)^2+(\epsilon_1^\prime-\epsilon_2^\prime)\epsilon_1^\dprime\leq 0$, which does not allow $\epsilon_1^\dprime$ to be small, and the corresponding bounds are irrelevant for our purpose.

Since, $\bfmu_1$ is a small isotropic perturbation of $\epsOne$ it is reasonable to assume that the effective Y-tensors also differ by a small perturbation denoted by $\bfdelta$, 
\begin{equation}\label{perturbed Y tensor relation 1}
    \bfY(\bfepseff) 
    =
    \bfY(\bfmu_*) + \bfdelta
\end{equation}
Using \eqref{perturbed Y tensor relation 1} in \eqref{CG relation 2}, and noting from our assumption that $\epsilon_2 = \mu_2$, we get,
\begin{equation}\label{CG relation 3}
    \bfY(\bfmu_*) + \bfdelta
    \leq 
     \bfY(\bfmu_*).
\end{equation}
On dividing \eqref{CG relation 3} by $\epsilon_1^\dprime$ and further simplifying the result gives, 
\begin{equation}\label{CG relation 4}
\begin{split}
    \frac{\bfdelta}{\epsilon_1^\dprime}
    \leq 
     0,
     \quad
     \implies
     -
    \frac{\bfY(\bfmu_*)-\bfY(\bfepseff)}{\epsilon_1^\dprime}
    \leq 
     0
\end{split}
\end{equation}
In the limit the perturbations become very small,  $\lim \epsilon_1^\dprime \to 0$, the above inequality gives us bounds on the sensitivity represented by the derivative $\partial \bfY(\bfepseff) / \partial \epsilon_1$:
\begin{equation}\label{sensitivity bounds Y-transform}
    \begin{split}
    \parderiv{\bfY(\bfepseff)}{\epsilon_1}
    \geq 
     0.
    \end{split}
\end{equation}

Now let the the perturbation $\epsilon_1^\dprime$ be purely imaginary (corresponding to low-loss composites). 
By the Cauchy-Riemann equations, the sensitivity bound in \eqref{sensitivity bounds Y-transform} then implies 
\begin{equation}
    \IM{(\bfY_*(\bfepseff))} = \IM{(\bfdelta)}\geq 0,
\end{equation}
which is implied by the wedge bound.
For the other bound, we assume $\mu_1=\epsilon_1$ in the first inequality of \eqref{CG relation 2} and proceed as before.
The main outcome here is that the bounds on the sensitivity implied  by the Cherkaev-Gibiansky bounds
\begin{equation}\label{CG bounds relation 2 again}
     \frac{\epsilon_1}{\mu_1} \bfY(\bfmu_*)
    \leq
    \bfY(\bfepseff)
    \leq 
    \frac{\epsilon_2}{\mu_2} \bfY(\bfmu_*),
\end{equation}
 are equivalent to the wedge bounds given in \eqref{lambda limit 1} and \eqref{lambda limit 2} when we take the low-loss limit.

It should be noted that, in addition to \eqref{CG relation 2}, when the constituent phases satisfy \eqref{phase case B}, Cherkaev and Gibiansky show that the effective tensors also satisfy another condition  given by, 
\begin{equation}\label{CG condition 2}
\begin{split}
      &\frac{\det{\left(\bfY(\bfepseff)/\epsilon_i-\bfR\bfY(\bfmu_*)^{-1}\bfR^T\mu_i\right)}}
    {(1-\det{(\bfY(\bfepseff)/\epsilon_i)})(1-\det^{-1}(\bfY(\bfmu_*)/\mu_i))}
    \leq
    \frac{(\epsilon_1\mu_1-\epsilon_2\mu_2)^2}{(\epsilon_1^2-\epsilon_2^2)(\mu_2^2-\mu_1^2)}, 
    \quad i=1,2,
\end{split}
\end{equation}
where
\begin{equation}
    \bfR = \begin{bmatrix}
        0&1\\-1&0
    \end{bmatrix}.
\end{equation}
As before, setting $\epsilon_1 = \epsilon_1^\prime$, $\mu_1 = \epsilon_1^\prime +\epsilon^\dprime$, $\mu_2 = \epsilon_2 = \epsilon_2^\prime$ and using the homogeneity of $\bfY(\bfepseff)$ we get
\begin{equation}
\begin{split}
      \frac{\det{\big(\bfY(\bfepseff/\epsilon_i) 
      -
      \bfR\bfY(\bfmu_*/\mu_i)^{-1}\bfR^T\big)} 
      \det{(\bfY(\bfmu_*/\mu_i))}}
    {(1-\det{(\bfY(\bfepseff/\epsilon_i))})
    (\det(\bfY(\bfmu_*/\mu_i))-1)}
    \leq
    \frac{(\epsilon_1^\prime(\epsilon_1^\prime+\epsilon_1^{\dprime})-{\epsilon_2^\prime}^2)^2}
    {({\epsilon_1^\prime}^2-{\epsilon_2^\prime}^2)({\epsilon_2^\prime}^2-(\epsilon_1^\prime+\epsilon_1^{\dprime})^2)}, 
\end{split}
\end{equation}
and using the relation $\det\big(\bfR\bfY\bfR^T\big) = \det{(\bfY)}$ gives
\begin{equation}\label{determinant inequality from CG}
    \begin{split}
    \frac{\det{\big(\bfY(\bfepseff/\epsilon_i)\bfR\bfY(\bfmu_*/\mu_i)^{-1}\bfR^T 
      -
      \bfI\big)} }
    {\det(\bfY(\bfmu_*/\mu_i))
    + \det(\bfY(\bfepseff/\epsilon_i))
    -\det(\bfY(\bfepseff/\epsilon_i))
    \det(\bfY(\bfmu_*/\mu_i))
    -1
    }
    \leq
    \frac{(\epsilon_1^\prime(\epsilon_1^\prime+\epsilon_1^{\dprime})-{\epsilon_2^\prime}^2)^2}
    {({\epsilon_1^\prime}^2-{\epsilon_2^\prime}^2)({\epsilon_2^\prime}^2-(\epsilon_1^\prime+\epsilon_1^{\dprime})^2)}.
    \end{split}
\end{equation}
For $i=1$, the inequality in \eqref{determinant inequality from CG} gives
\begin{equation}\label{determinant CG 1}
\begin{split}
&\frac{\det{\big(\bfY(\bfepseff/\epsilon_1^\prime)\bfR\bfY(\bfmu_*/(\epsilon_1^\prime+\epsilon_1^\dprime))^{-1}\bfR^T 
      -
      \bfI\big)} }
    {\det(\bfY(\bfmu_*/(\epsilon_1^\prime+\epsilon_1^\dprime)))
    + \det(\bfY(\bfepseff/\epsilon_1^\prime))
    -\det(\bfY(\bfepseff/\epsilon_1^\prime))
    \det(\bfY(\bfmu_*/(\epsilon_1^\prime+\epsilon_1^\dprime)))
    -1
    }
    \\
    &\leq
    \frac{(\epsilon_1^\prime(\epsilon_1^\prime+\epsilon_1^{\dprime})-{\epsilon_2^\prime}^2)^2}
    {({\epsilon_1^\prime}^2-{\epsilon_2^\prime}^2)({\epsilon_2^\prime}^2-(\epsilon_1^\prime+\epsilon_1^{\dprime})^2)}, 
\end{split}
\end{equation}
and for $i=2$ it gives
\begin{equation}\label{determinant CG 2}
\begin{split}
&\frac{\det{\big(\bfY(\bfepseff/\epsilon_2^\prime)\bfR
\bfY(\bfmu_*/\epsilon_2^\prime)^{-1}
\bfR^T 
      -
      \bfI\big)} }
    {\det(\bfY(\bfmu_*/\epsilon_2^\prime))
    + \det(\bfY(\bfepseff/\epsilon_2^\prime))
    -\det(\bfY(\bfepseff/\epsilon_2^\prime))
    \det(\bfY(\bfmu_*/\epsilon_2^\prime))
    -1
    }
   \leq
    \frac{(\epsilon_1^\prime(\epsilon_1^\prime+\epsilon_1^{\dprime})-{\epsilon_2^\prime}^2)^2}
    {({\epsilon_1^\prime}^2-{\epsilon_2^\prime}^2)({\epsilon_2^\prime}^2-(\epsilon_1^\prime+\epsilon_1^{\dprime})^2)}.
\end{split}
\end{equation}
The left hand sides of 
 \eqref{determinant CG 1} and \eqref{determinant CG 2} approach zero as $\epsilon_1^\dprime \to 0$, while the right hand sides do not. 
 Therefore, these inequalities are irrelevant in the small $\epsilon_1^\dprime$ limit.
Hence, the condition in \eqref{CG condition 2} does not contribute to the sensitivity bounds.
As the Cherkaev-Gibiansky bounds are optimal, completely characterizing the G-closure, the associated sensitivity bounds implied by \eqref{CG bounds relation 2 again} also completely characterize the possible sensitivities for each $\bfepseff$, and will be attained by the same microgeometries that attain the Cherkaev-Gibiansky bounds.
Hence, the wedge bounds should at least almost characterize the possible complex $\bfepseff$ in the low-loss limit.




\section{Conclusion}
In this work, we systematically investigated the bounds and optimal microstructures associated with the complex uniaxial effective permittivity of two-phase dielectric composites in the quasistatic regime. We demonstrated that the classical bounds on the transverse component $\epsilon_\perp$ of the uniaxial effective permittivity, given by \cite{milton1981bounds}, are not optimal. Through extensive numerical analysis of hierarchical laminates (HLs), we conjecture improved bounds for $\epsilon_\perp$ in the form of circular arcs, supported by convex hull computations over millions of generated microstructures. 
Gaps in attainability of the region 
$\Psi$ by HLs are perhaps to be expected since the analogous arc to $MCN$ (with
$\parderivsec{\epsilon_\perp}{\epsilon_1}\Big\vert_{\epsilon_1=1}=-f_1f_2$) in the isotropic case (with $\parderivsec{\epsilon_*}{\epsilon_1}\Big\vert_{\epsilon_1=1}=-f_1f_2/3$) also appears to have such gaps \citep{kern2020tight}.

We have identified two rank-4 HL families that attain all points on the conjectured bounds.
Furthermore, we analyzed the correlation between $\epsilon_\perp$ and $\epsilon_\parallel$ and developed a design algorithm that allows the construction of HL geometries achieving a specified transverse component $\epsilon_\perp$. We find that the correlated axial components for transverse components on the boundaries $AM$ and $AN$ of the region $\Psi$ are restricted to the boundary of $\Omega^\prime$.
Conversely, the correlated transverse components for axial components on the boundary of $\Omega^\prime$ lie within the region $\Psi$ as well.

Using the Cherkaev-Gibiansky transformation and the method of translations, we extend the tight bounds developed for isotropic composites by \cite{kern2020tight} to the uniaxial case. 
While the translation-based bounds recover the known isotropic results in the  limit $\epsilon_\perp\to\epsilon_\parallel$, we find that in the uniaxial case when we seek bounds on $\epsilon_\perp$ that do not involve $\epsilon_\parallel$ (or vice-versa), the specific form of isotropic translation tensor employed does not yield improved bounds over those originally given by \cite{milton1981bounds}.
For anisotropic two-dimensional two-phase composites with low loss, optimal bounds on the sensitivity of the anisotropic effective permittivity with respect to the constituent permittivity $\epsilon_1$ are obtained. The sensitivity bounds imply the optimality of the wedge bounds given by \cite{milton1987multicomponent} in the low loss limit.


Several interesting directions remain open for further investigation. 
First, a rigorous analytical proof of the conjectured optimal bounds on $\epsilon_\perp$ remains an open problem.
Second, the role of anisotropic translation tensors in deriving tighter theoretical bounds on uniaxial effective permittivity deserves systematic study, extending the work of \cite{kern2020tight} to fully anisotropic cases. 
Third, it would be interesting to see if the conjectured optimal HLs can be replaced by more realistic structures, perhaps using topology optimization.
Fourth, given the relevance of uniaxial composites to hyperbolic metamaterials, an important avenue for future work is to apply these design principles to the realization of optimized HMMs with tailored dispersion properties for applications in sub-diffraction imaging, structures for optical negative refraction, spontaneous emission control, and other photonic devices.
Finally, it would be good if for two-dimensional, two-phase composites the set of all possible complex effective permittivity tensors could be found and the microgeometries identified  for arbitrary $\epsilon_1$ and $\epsilon_2$, and not just in the low loss limit.

We hope that the results and methods presented in this work serve as theoretical and  computational design tools to guide future explorations in the development of optimized uniaxial composites and advanced hyperbolic metamaterials.

\section*{Acknowledgement}

The authors are grateful to the National Science Foundation for support through grant DMS-2107926, and to Christian Kern for allowing us to use his results (Figure \ref{fig:Point C locus}) and for many helpful comments on the manuscript.








\bibliographystyle{mod-xchicago}
\bibliography{main}

\end{document}